\documentclass{article}
\title{Search for gauge symmetry generators of singular Lagrangian theory.}
\author{A.A. Deriglazov\footnote{alexei@ice.ufjf.br ~ On leave of
absence from Dept. Math. Phys., Tomsk Polytechnical University,
Tomsk, Russia.}}
\date{Dept. de Matematica, ICE, Universidade Federal de Juiz de Fora,\\
MG, Brazil.}
\begin{document}
\maketitle
\large

\begin{abstract}
We propose a procedure which allows one to construct local symmetry generators of general quadratic  
Lagrangian theory. Manifest recurrence relations for 
generators in terms of so-called structure matrices of the Dirac formalism are obtained. The procedure 
fulfilled in terms of initial variables of the theory, and do not implies either separation of constraints 
on first and second class subsets or any other choice of basis for constraints.
\end{abstract}

\noindent

\section{Introduction}

Relativistic theories are usually formulated in manifestly covariant form, i.e. in the form with linearly 
realized Lorentz group. It is achieved by using of some auxiliary variables, which implies appearance of 
local (gauge) symmetries in the corresponding Lagrangian action. Investigation of the symmetries is 
essential part of analysis of both classical and quantum versions of a theory. 
Starting from pioneer works on canonical quantization of singular theories [1-3], 
one of the most intriguing problems is search for constructive procedure 
which allows one to find the local symmetries from known Lagrangian or Hamiltonian 
formulation [4-10]. For a theory with first class constraints only, such a kind procedure has been proposed 
in [4, 5]. Symmetry structure of a general singular theory has been described in recent  
works [6-8]. In particular, it was shown how one can find gauge symmetries of Hamiltonian action for 
quadratic theory [7], as well as for general singular theory [8]. 

In the present work we propose an alternative procedure to construct Lagrangian local symmetries 
for the case of general quadratic theory. Our method is based on analysis of Noether identities in the 
Hamiltonian form, the latter has been obtained in our works [9, 10]. 
Let us enumerate some characteristic properties of the procedure presented below. 

\noindent
1) The procedure do not implies separation of Hamiltonian constraints on first and second class 
subsets, which is may be the most surprising result of the work.

\noindent
2) The procedure do not requires choice of some special basis for constraints. 

\noindent
3) All the analysis is fulfilled in terms of initial variables of a theory. 

Rather schematically, final result of our work can be described as follows. Let
$H=H_0+\Phi_\alpha v^{\alpha}$ be Hamiltonian of a theory, 
where $v^{\alpha}$ denotes Lagrangian multipliers to all primary constraints $\Phi_\alpha$.
Let $\Phi^{(s-1)}{}_{\mu}$ 
be constraints of $(s-1)$-stage of the Dirac procedure. On the next stage one studies the equations
$\{\Phi^{(s-1)}{}_{\mu}, H\}=0$  
for revealing of $s$-stage constraints. $s$-stage Dirac functions 
$\{\Phi^{(s-1)}{}_{\mu}, H\}$ can be rewritten in the form  
\begin{eqnarray}\label{101}
\{\Phi^{s-1}_{\mu}, H\}=
A_{\mu}{}^{\nu} 
\left(
\begin{array}{c}
\pi^{(s)}{}_i(v)\\
\Phi^{(s)}{}_{\zeta}+B_{\zeta}(\pi^{(s-1)}, \ldots , \pi^{(2)})\\
C_a(\Phi^{s-1}, \ldots ,\Phi^2)+
D_a(\pi^{(s-1)}, \ldots ,\pi^2),
\end{array}
\right)
\end{eqnarray}
where 
$\pi^{(s)}{}_i(v)=0$ represent equations for determining of the Lagrangian multipliers of these stage,  
$\Phi^{(s)}{}_{\zeta}$ are $s$-stage constraints, and $A, B, C, D$ will be called $s$-stage structure 
matrices. It may happens that some components of the column 
(\ref{101}) do not represent independent restrictions on the variables $(q, p, v)$ of the theory. 
The number $[a]$ of these components will be called defect of $s$-stage system. Then  
$[a]$ independent local symmetries of the Lagrangian action can be constructed   
\begin{eqnarray}\label{102}
\delta q^A=
\sum^{s-2}_{p=0}{\stackrel{(p)}{\epsilon}{}^a}
R^{(p)}_a{}^{A}(q, \dot q),
\end{eqnarray}
where generators $R$ are specified in terms of the structure matrices in an algebraic way. 
Recurrence relations for 
obtaining of the generators are presented below (see Eqs.(\ref{901})-(\ref{903}) for s-stage symmetries and 
Eqs.(\ref{906})-(\ref{908}) for lower-stage symmetries).
 
Thus, knowledge of a structure of $s$-stage Dirac functions (\ref{101}) is equivalent to knowledge 
of $s$-stage local symmetries (\ref{102}).
Total number of independent symmetries, which can be find by using of our procedure,   
coincides with the number of Lagrangian multipliers remaining undetermined in the Dirac procedure. 

\section{Setting up}

We consider Lagrangian theory with action being 
$(A=1,2,\cdots [A])$
\begin{eqnarray}\label{201}
S=\int d\tau L(q^A,\dot q^A), \qquad 
\end{eqnarray}
It is supposed that the theory is singular  
\begin{eqnarray}\label{202}
rank\frac{\partial^2 L}{\partial\dot q^A\partial\dot q^B}= [i]<[A]. 
\end{eqnarray}
According to Dirac [1], Hamiltonian formulation of the theory is obtained as follow. 
First stage of Hamiltonization procedure is to define equations for   
the momenta $p_A$: $ ~ p_A=\frac{\partial L}{\partial \dot q^A} ~$. Being     
considered as algebraic equations for determining 
of velocities $\dot q^A$, $[i]$ equations 
can be resolved for $\dot q^i$ and then substituted into the remaining ones. By construction, the 
resulting equations do not depend on $\dot q^A$ and are called primary constraints $\Phi_\alpha(q, p)$ 
of the Hamiltonian formulation. The equations $p_A=\frac{\partial\bar L}{\partial \dot q^A}$ 
are then equivalent to the following system  
\begin{eqnarray}\label{203}
\dot q^i=v^i(q^A, p^i, \dot q^\alpha),
\end{eqnarray}
\begin{eqnarray}\label{204}
\Phi_\alpha\equiv p_{\alpha}-f_{\alpha}(q^A, p_j)=0.
\end{eqnarray}
Then one introduces an extended phase space with the coordinates $(q^A, p_A, v_\alpha)$. 
By definition, Hamiltonian formulation of the theory (\ref{201}) is the following system of 
equations on this space 
\begin{eqnarray}\label{205}
\dot q^A=\{q^A, H\}, ~  \dot p_A=\{p_A, H\}, \cr 
\Phi_\alpha(q^A, p_B)=0, \qquad
\end{eqnarray}
where $\{ ~ , ~ \}$ is the Poisson bracket, and it was denoted 
\begin{eqnarray}\label{206}
H(q^A, p_A, v^\alpha)=H_0(q^A, p_j)+v^\alpha\Phi_\alpha(q^A, p_B),
\end{eqnarray}
\begin{eqnarray}\label{207}
H_0=\left(p_i\dot q^i-L+ \dot q^\alpha\frac{\partial\bar 
L}{\partial \dot q^\alpha}\right) 
\Biggr|_{\dot q^i\rightarrow v^i(q^A, p_j, \dot q^\alpha)}.
\end{eqnarray}
The variables $v^{\alpha}$ are called Lagrangian multipliers to the primary constraints. 
It is known [11] that the formulations (\ref{201}) and (\ref{205}) are equivalent. 

Second stage of the Dirac procedure consist in analysis of second stage equations 
$\{\Phi_{\alpha}, H\}=0$, the latter are algebraic consequences of the system (\ref{205}).
Some of second-stage equations can be used for determining of a subgroup of Lagrangian multipliers 
in an algebraic way. Among the remaining equations one takes functionally independent subsystem, the latter 
represent secondary Dirac constraints $\Phi^{(2)}_{\alpha_2}(q^A, p_j)=0$. They imply third-stage equations, 
an so on. We suppose that the theory has 
constraints up to at most $N$ stage: $(\Phi_{\alpha}, \Phi^{(2)}_{\alpha_2}, \ldots , \Phi^{(N)}_{\alpha_N})$. 
  
\section{Sufficient conditions for existence of local symmetry of Lagrangian action}

Let us consider infinitesimal transformation 
\begin{eqnarray}\label{301}
q^A\longrightarrow q^{'A}=q^A+\delta q^A, \qquad
\delta q^A=
\sum^{[p]}_{p=0}{\stackrel{(p)}{\epsilon}}
R^{(p)}{}^A(q, \dot q, \ddot q, ...),
\end{eqnarray}
where {\sf parameter} $\epsilon(\tau)$ is arbitrary function of time $\tau$, 
and it was denoted
$\stackrel{(p)}{\epsilon}\equiv\frac{d^p}{d\tau^p}\epsilon$. The transformation is 
{\sf local (or gauge) symmetry}  
of an action $S$, if it leaves $S$ invariant up to surface term
\footnote{Transformations which involve variation of the evolution parameter: $\tilde\delta\tau,
\tilde\delta q^A$ are included into the scheme. Actually, with any such transformation it is associated 
unambiguously transformation of the form (\ref{301}): $\delta\tau=0$,
$\delta q^A=-\dot q^A\tilde\delta\tau+\tilde\delta q^A$. If $\tilde\delta q$
is a symmetry of the action, the same will be true for $\delta q$.}
\begin{eqnarray}\label{302}
\delta L=\frac{d}{d\tau}\omega,
\end{eqnarray}
with some functions $\omega(q, \epsilon)$. Operators $\sum^{[p]}_{p=0}R^{(p)}{}^A\frac{d^p}{d\tau^p}$ 
will be called {\sf generators} of the local symmetry. Local symmetry implies appearance of 
identities among equations of motion of the theory. For a theory without higher derivatives and 
generators of the 
form\footnote{Analysis of this work shows that gauge generators can be find in this form.} 
$R^{(p){}A}(q, \dot q)$, the identities were analyzed in some 
details in our  
recent work [10]. Hamiltonian form of the identities has been obtained, then necessary and sufficient 
conditions for existence of local symmetry of the Lagrangian action were formulated on this ground. 
Namely, the Hamiltonian identities can be considered as a system of partial differential equations  
for the Hamiltonian counterparts of the functions $R^{(p)}{}^A(q, \dot q)$. In the present work 
we propose pure algebraic procedure to solve these equations. So, let us present 
the relevant result of the work [10] in a form convenient for subsequent analysis. 

For given integer number $s$, let us construct {\sf generating functions} $T^{(p)}, ~ p=2,3, \ldots ,s$ 
according to the recurrence relations ($T^{(1)}=0$)
\begin{eqnarray}\label{303}
T^{(p)}=Q^{(p) \alpha}\{\Phi_\alpha, H\}+\{H, T^{(p-1)}\},
\end{eqnarray}
where the coefficients $Q^{(p) \alpha}(q^A, p_j, v^\alpha)$ are some functions. Then one can prove [10] 
the following\footnote{ Equations (\ref{304}), (\ref{305}) has been obtained in [10] starting from 
hypothesis that the action is invariant, and by making substitution $v^i(q^A, p_j, v^\alpha)$ into the 
first order identities, i. e. as necessary conditions for existence of gauge symmetry. 
As it is explained in [12], this substitution is change of variables 
on configuration-velocity space, which implies that (\ref{304}), (\ref{305}) are sufficient conditions also.}

{\bf{Statement 1.}} Let the coefficients $Q^{(p) \alpha}, p=2,3, \ldots ,s$ have been chosen in such a way that 
the following equations:
\begin{eqnarray}\label{304}
\frac{\partial}{\partial v^\alpha}T^{(p)}=0, \quad
p=2,3 \ldots ,s-1,
\end{eqnarray}
\begin{eqnarray}\label{305}
T^{(s)}=0, \qquad \qquad \qquad \qquad
\end{eqnarray}
hold. Using these $Q$, let us construct the Hamiltonian functions $R^{(p) A}(q^A, p_j, v^\alpha)$, 
$p=0,1,2, \ldots ,s-2$
\begin{eqnarray}\label{306}
R^{(p) \alpha}=Q^{(s-p) \alpha}, \qquad
R^{(p) i}=\{q^i, \Phi_\alpha\} R^{(p)\alpha}-\{q^i, T^{(s-1-p)}\}, 
\end{eqnarray}
and then the Lagrangian functions
\begin{eqnarray}\label{307}
R^{(p) A}(q, \dot q)\equiv
R^{(p) A}(q^A, p_j, v^\alpha)\Biggr|_{p_j\rightarrow\frac{\partial\bar L}{\partial v^j}}
\Biggr|_{v^A\rightarrow\dot q^A}.
\end{eqnarray}
Then the transformation 
\begin{eqnarray}\label{308}
\delta q^A=
\sum^{s-2}_{p=0}{\stackrel{(p)}{\epsilon}}
R^{(p) A}(q, \dot q),
\end{eqnarray}
is local symmetry of the Lagrangian action.

Search for the symmetry (\ref{308}) (the latter involves derivatives of the parameter 
$\epsilon$ up to order $s-2$)  
is directly related with $s$-stage of the Dirac procedure, see below. On this reason the symmetry 
(\ref{308}) will be called {\sf $s$-stage symmetry}. Then the set $T^{(2)}, T^{(3)}, \ldots ,T^{(s)}$ 
can be called {\sf $s$-stage generating functions}. 
Some relevant comments are in order.

1) From Eq.(\ref{306}) it follows that only $R^{(p) \alpha}$-block of Hamiltonian generators is 
essential quantity. On this reason, only this block will be discussed below.

2) Hamiltonian functions (\ref{306}) can be used also to construct a local symmetry of the Hamiltonian 
action. Expressions for the corresponding transformations $\delta q^A, \delta p_A, \delta v^\alpha$ can be 
find in [10].

3) According to the statement, symmetries of different stages $s$ can be looked for separately. 
To find 2-stage 
symmetries $\delta_{a_2} q^A=\epsilon^{a_2} R_{a_2}^{(0) A}$, one look for solutions $Q_{a_2}^{(2) \alpha}$ 
of the 
equation 
\begin{eqnarray}\label{309}
T^{(2)}\equiv Q^{(2) \alpha}\{\Phi_\alpha, H\}=0.
\end{eqnarray}
Note that it implies analysis of second-stage Dirac functions $\{\Phi_\alpha, H\}$. 
3-stage symmetries 
$\delta_{a_3} q^A=\epsilon^{a_3} R_{a_3}^{(0) A}+\dot\epsilon^{a_3} R_{a_3}^{(1) A}$
are constructed from solutions $Q_{a_3}^{(2) \alpha}, Q_{a_3}^{(3) \alpha}$ 
of the equations{\footnote{Coefficients $Q$ of different stages are independent.}} 
\begin{eqnarray}\label{310}
\frac{\partial}{\partial v^\beta}
T^{(2)}\equiv \frac{\partial}{\partial v^\beta}(Q^{(2) \alpha}\{\Phi_\alpha, H\})=0, \cr
T^{(3)}\equiv Q^{(3) \alpha}\{\Phi_\alpha, H\}+
\left\{Q^{(2)\alpha}\{\Phi_\alpha, H\}, H\right\}=0,
\end{eqnarray}
and so on. In a theory with at most $N$-stage Dirac constraints presented, the procedure stops for 
$s=N+1$, see Sect. 9 below. 

4) {\sf Generating equations} (\ref{304}), (\ref{305}), are related with $s$-stage of 
the Dirac procedure in the following sense. 
In Sect. 6 we demonstrate that the coefficients $Q^{(p)},~ p=2, 3, \ldots ,s$ 
can be chosen in such a way that each generating function $T^{(p)}$ is linear combination of constraints 
$\Phi_{\alpha_k}$ of the stages $k=2, 3, \ldots ,p$. In particular, 
$T^{(s-1)}=\sum^{s-1}_{p=2}{c^{\alpha_p}\Phi_{\alpha_p}}$, then 
$T^{(s)}\sim\{T^{(s-1)}, H\}$ in Eq.(\ref{305})
involves the Dirac functions up to $s$-stage: $\{\Phi_{\alpha_p}, H\},~ p=2, 3, \ldots ,s-1$.  
So, search for $s$-stage symmetries implies analysis of $s$-stage of the Dirac procedure.  

5) Since the generating equations (\ref{304}), (\ref{305}) do not involve the momenta $p_\alpha$, one can 
search for solutions in the form $Q^{(p) \alpha}(q^A, p_j, v^\alpha)$. As a result, Hamiltonian generators 
do not depend on $p_\alpha$. In this case, passage to the Lagrangian first order formulation 
is change of variables [12]: $(q^A, p_i, v^\alpha)\leftrightarrow(q^A, v^i, v^\alpha)$. 
This change has been performed in Eq. (\ref{307}). 

6) Analysis of the Hamiltonian equations (\ref{304}), (\ref{305}) turns out to be more simple task 
as compare to the corresponding Lagrangian (or first order) version. Besides the fact that one is 
able to use well 
developed Dirac method, crucial simplification is linearity on $v^\alpha$ of all the Hamiltonian 
quantities appeared in the analysis. On this reason, search for solutions of the generating  
equations can be reduced to solving of some system of linear inhomogeneous equations, see 
below. Lagrangian version of our procedure will imply looking for solutions of some non linear 
algebraic equations on each step of the procedure.   

\section{Analysis of second-stage Dirac functions}

As it was discussed in the previous section, expression for generating function $T^{(k)}$ involve the  
Dirac functions $\{\Phi_{\alpha_p}, H\}$, $p=1, 2, \ldots ,k-1$. One needs to know detailed 
structure of them to solve the generating equations. Let us point that 
this part of analysis is, in fact, part of the Dirac procedure for revealing of higher-stage 
constraints. The only difference is that in the Dirac procedure one studies the equations   
$\{\Phi_{\alpha_p}, H\}=0$, where constraints and equations for Lagrangian multipliers 
of previous stages can be used. Since our generating equations must be satisfied by $Q$ for 
any $q, p, v$, 
one needs now to study the Dirac functions outside of extremal surface. Below we suppose that 
matrices $\{\Phi_{\alpha_p}, \Phi_{\alpha_1}\}$ have constant rank in vicinity of phase space point 
under consideration. In particular, it is true for quadratic theory.
In the next section we formulate an induction 
procedure to represent $p$-stage Dirac functions in {\sf normal form} convenient for 
subsequent analysis, see Eq.(\ref{601}) below. On each stage, it will be necessary to 
divide some groups of functions on subgroups. Here we present detailed analysis of second stage, with 
the aim to clarify notations which will be necessary to work out $p$-stage Dirac functions and the 
corresponding generating equations.

With a group of quantities appeared on first stage of the Dirac procedure we assign number 
of the stage, the latter replace corresponding index (the number will be called 
{\sf index of the group} below).  
Then the primary constraints are $\Phi_\alpha\equiv\Phi_1$, and the Lagrangian 
multipliers are denoted as $v^{\alpha}\equiv v^1$. Number of functions in a group is denoted as 
$[1]\equiv [\alpha]$.  
For the second stage Dirac functions one writes
\begin{eqnarray}\label{401}
\{\Phi_\alpha, \Phi_\beta v^\beta +H_0\} \longrightarrow
\{\Phi_1, \Phi_{1'} v^{1'} +H_0\}= \cr 
\{\Phi_1, \Phi_{1'}\} v^{1'} +\{\Phi_1, H_0\}\equiv
\triangle_{(2)1 1'}v^{1'}+H_{(2)1}.
\end{eqnarray}
So, repeated up and down number of stage imply summation over the corresponding indices.
With quantities first appeared on second stage has been assigned number of the stage: 
$\triangle_{(2)}, ~ H_{(2)}$ 
(where confusion is not possible, it can be omitted).  

Let us describe a procedure to represent second-stage Dirac functions (\ref{401}) in the 
normal form. Suppose that $rank\triangle_{(2)1 1'}=\left[\overline{2}\right]$, then one finds 
$\left[\widetilde{2}\right]=[1]-\left[\overline{2}\right]$ 
independent null-vectors $\vec K_{(2) \widetilde{2}}$ of the matrix $\triangle_{(2)}$ with components 
$K_{(2) \widetilde{2}}{}^1$. Let $K_{(2) \overline{2}}{}^1$ be any completion of the set 
$\vec K_{(2) \widetilde{2}}$ up to a basis of $[1]$-dimensional space. By construction, the matrix    
\begin{eqnarray}\label{402}
K_{(2) \widehat 1}{}^1\equiv
\left(K_{(2) \overline{2}}{}^1\atop  K_{(2) \widetilde{2}}{}^1\right),
\end{eqnarray}
is invertible, with the inverse matrix being 
$\widetilde K_{(2)1}{}^{\widehat 1} K_{(2) \widehat 1}{}^{1'}=\delta_1{}^{1'}$. 
The matrix $K$ is a kind of {\sf conversion matrix} which transforms the index $1$ into  
$\widehat 1$, the latter is naturally  divided on two groups 
\begin{eqnarray}\label{403}
1\rightarrow \widehat 1=(\bar 2, \widetilde{2}).
\end{eqnarray}
Since $K_{(2) \widetilde{2}}{}^1\triangle_{(2)1 1'}=0$, 
the conversion matrix can be used to separate the Dirac functions on $v$-dependent and $v$-independent 
parts: $\{\Phi_1, H\}=\widetilde K K \{\Phi_1, H\}$, with the result being
\begin{eqnarray}\label{404}
\{\Phi_1, H\}=
\widetilde K_{(2) 1}{}^{\widehat 1}\left(\pi_{\overline{2}}(v^1)\atop 
\Phi_{\widetilde{2}}(q^A, p_j)\right), \\[1pt] \nonumber
\pi_{\overline{2}}(v^1)\equiv X_{(2)\overline 2 1}v^1+Y_{(2) \overline 2}, \cr 
\Phi_{\widetilde{2}}\equiv K_{(2) \widetilde{2}}{}^1 H_{(2)1}.
\end{eqnarray}
Here it was denoted 
\begin{eqnarray}\label{405}
X_{(2)\overline 2 1}=K_{(2) \overline 2}{}^{1'}\triangle_{(2)1' 1}, \qquad  
Y_{(2) \overline 2}=K_{(2) \overline 2}{}^1 H_{(2)1}.
\end{eqnarray}
Let us analyze the functions $\pi_{\overline{2}}(v^1)$. By construction, the matrix $X$ has maximum rank 
equal $\left[\overline{2}\right]$. Without loss of generality, we suppose that from the beginning $v^1$ 
has been chosen 
such that the rank columns appear on the left: 
$X_{(2)\overline{2} 1}=(X_{(2)\overline{2} \overline{2}}, X_{(2)\overline{2} \underline{2}})$. 
So, the Lagrangian multipliers are divided on two groups 
$v^1=(v^{\overline{2}}, v^{\underline{2}})$, one 
writes\footnote{On this stage one has $[\widetilde{2}]=[\underline{2}]$, but it will not 
be true for higher stages. On this reason we adopt different notations for these groups.}
\begin{eqnarray}\label{406}
\pi_{\overline{2}}=X_{(2) \overline{2} \overline{2}}v^{\overline{2}}+
X_{(2) \overline{2} \underline{2}}v^{\underline{2}}
+Y_{(2) \overline{2}}.
\end{eqnarray}
Then $v^{\bar 2}$ can be identically rewritten in terms of 
$v^{\underline{2}}, \pi_{\overline{2}}$ as follows:
\begin{eqnarray}\label{407}
v^{\overline{2}}\equiv\widetilde X_{(2)}{}^{\overline{2} \overline{2'}}\pi_{\overline{2'}}(v^1)+
\Lambda_{(2)}{}^{\overline{2}}{}_{\underline{2}}v^{\underline{2}}+W_{(2)}{}^{\overline{2}},
\end{eqnarray}
where
\begin{eqnarray}\label{4077}
\Lambda_{(2)}{}^{\overline{2}}{}_{\underline{2}}=-\widetilde X_{(2)}{}^{\overline{2} \overline{2'}}
X_{(2) \overline{2'} \underline{2}}, \qquad 
W_{(2)}{}^{\overline{2}}=-\widetilde X_{(2)}{}^{\overline{2} \overline{2'}}Y_{(2) \overline{2'}}. 
\end{eqnarray}
We stress that Eq.(\ref{407}) is an identity. It will be necessary to analyze third-stage Dirac 
functions below. 

Let us analyze the functions $\Phi_{\widetilde{2}}$ in Eq.(\ref{404}). By construction, they depend 
on the phase space variables $z_1\equiv(q^A, p_j)$. According to Dirac, functionally independent functions 
among $\Phi_{\widetilde{2}}$ are called secondary constraints, and the equations $\Phi_{\widetilde{2}}=0$ 
can be used to express a part $\bar z_2$ of the phase space variables $z_1=(\bar z_2, z_2)$  
in terms of $z_2$. Let us suppose 
\begin{eqnarray}\label{408}
rank\frac{\partial\Phi_{\widetilde{2}}}{\partial z_1}\Biggr|_{\Phi_{\widetilde{2}}}=
rank\frac{\partial\Phi_{\widetilde{2}}}{\partial \bar z_2}\Biggr|_{\Phi_{\widetilde{2}}}=[\bar z_2].
\end{eqnarray}
We demonstrate that the functions $\Phi_{\widetilde{2}}$ can be identically rewritten in the form  
\begin{eqnarray}\label{409}
\Phi_{\widetilde{2}}(z_1)=U_{(2) \widetilde{2}}{}^{\widetilde{2'}}
\left(\Phi_2(z_1)\atop 0_{\breve 2}\right), \qquad 
\left[\breve 2\right]=\left[\widetilde{2}\right]-\left[2\right], \qquad \det U\ne 0,
\end{eqnarray}
where index $\widetilde 2$ is divided on two groups $\widetilde 2=(2, \breve 2)$, and $\Phi_2$ 
are functionally independent.

Actually, under the conditions (\ref{408}) there is 
exist\footnote{Let us point that only on this place of the Dirac procedure some nonlinear 
algebraic equations appear. As it will be shown below, analysis of the generating equations implies 
solution of linear systems only. In this sense, our procedure can be applied to any theory where 
the Dirac procedure works.} 
the representation [11] 
(see also Appendix A)
\begin{eqnarray}\label{410}
\Phi_{\widetilde{2}}=\Lambda_{(\widetilde{2})\times (2)}\Phi_2(z_1), \quad 
rank \Lambda=[2], \quad 
rank\frac{\partial\Phi_2}{\partial\bar z_2}\Biggr|_{\Phi_2}=[\Phi_2]=[\bar z_2].
\end{eqnarray}
Using invertible numerical matrix $Q$, lines of the matrix $\Lambda$ can be rearranged 
\begin{eqnarray}\label{411}
\Phi_{\widetilde{2}}=Q_{(\widetilde{2})\times(\widetilde{2})}
\left(\Lambda'_{(2)\times(2)}\atop \Lambda''_{(\breve 2)\times(2)}\right)\Phi_2, 
\end{eqnarray}
in such a way that $\det\Lambda'\ne 0$. Then one writes identically
\begin{eqnarray}\label{412}
\Phi_{\widetilde{2}}=Q
\left(
\begin{array}{cc}
\Lambda'&0 \\
\Lambda''&1
\end{array}
\right)
\left(\Phi_2\atop 0\right)\equiv U_{(2)}\left(\Phi_2\atop 0\right),
\end{eqnarray}
where $U_{(2)}$the invertible matrix.

Substitution of this result into Eq.(\ref{404}) gives the normal form of second-stage Dirac functions
\begin{eqnarray}\label{413}
\{\Phi_1, H\}=
A_{(2) 1}{}^{\widehat 1}(q^A, p_j)
\left(
\begin{array}{c}
\pi_{\overline{2}}(v^1)\\
\Phi_2(q^A, p_j)\\
0_{\breve 2}
\end{array}
\right)
\end{eqnarray}
where $A$ is invertible matrix  
\begin{eqnarray}\label{414}
A_{(2) 1}{}^{\widehat 1}=\widetilde K_{(2)}
\left(
\begin{array}{cc}
1_{(\overline{2})\times(\overline{2})}&0\\
0&U_{(2) \widetilde{2}}{}^{\widetilde{2}}
\end{array}
\right)
\end{eqnarray}
and functions $\pi_{\overline{2}}(v^1)$ are given by Eq.(\ref{406}). In the process, the Lagrangian 
multipliers $v^1$ have been divided on subgroups $(v^{\overline 2}, v^{\underline 2})$, where 
$v^{\overline 2}$ can be identically rewritten in terms of $v^{\underline 2}, \pi_{\underline 2}$ 
according to Eq.(\ref{407}). The functions 
$\Phi_\alpha=p_\alpha-f_\alpha(q^A, p_j), ~ \Phi_2(q^A, p_j)$ are functionally independent, 
$\Phi_2$ represent all secondary constraints of the theory. By construction, 
$\pi_{\overline 2}(v^1)=0$ turn out to be 
equations for determining of the Lagrangian multipliers $v^{\overline 2}$.

Note that "evolution" of the index $1$ of previous stage during the second stage can be 
resumed as follow: 
it can either be divided on two subgroups: $1=(\overline{2}, \underline{2})$, 
or can be converted into $\widehat 1$ and then divided on three subgroups:
$1\rightarrow\widehat 1=(\overline{2}, 2, \breve 2)$. Dimensions of the indices are related with 
rank properties of second stage Dirac system as follow:\par
\noindent
$\left[\overline{2}\right]$ is number of Lagrangian multipliers which can be determined on 
the second stage;\par
\noindent
$\left[\underline{2}\right]$ is number of multipliers which remains undetermined after 
the second stage;\par  
\noindent
$[2]$ is number of secondary constraints;\par 
\noindent
$\left[\breve 2\right]=\left[1\right]-\left[\overline{2}\right]-[2]$ is called 
{\sf defect of the system} (\ref{413}). 

\section{Notations}

Discussion of previous section on second stage of the Dirac procedure excuses the 
following notations.\par 

a) Notations for phase space variables ($p_{\alpha}$ are not included): $(q^A, p_j)\equiv z_1$. 
On second stage of the Dirac procedure the group 
can be divided on two subgroups $z_1=(\bar z_2, z_2)$, where $\bar z_2$ are variables which can be 
presented through $z_2$ using the secondary constraints. On third stage one has  
$z_2=(\bar z_3, z_3)$, and so on. In the end of Dirac procedure one obtains 
the following division: $z_1=(\bar z_2, \bar z_3, \ldots ,\bar z_N, z_N)$.\par

b) For the Lagrangian multipliers to the primary constraints we assign "covariant" index                                                                                                                                                                                                                                            
$v^{\alpha}\equiv v^1$. On second stage of the Dirac procedure the group 
can be divided on two subgroups , $v^1=(v^{\overline{2}}, v^{\underline{2}})$,     
where $v^{\overline{2}}$ represents subgroup which can be 
presented through $v^{\underline{2}}$ on this stage. On the next stages one has  
$v^{\underline{2}}=(v^{\overline{3}}, v^{\underline{3}}), \ldots , 
v^{\underline{N}}=(v^{\overline{N+1}}, v^{\underline{N+1}})$.
Thus the symbols $v^1, ~ v^{\underline{k}}, ~ 1<k<p-1$ appeared on the stage $p$ mean: 
\begin{eqnarray}\label{501}
v^{\underline{k}}=(v^{\overline{k+1}}, v^{\overline{k+2}}, 
\ldots ,v^{\overline{p-1}}, v^{\underline{p-1}}),
\end{eqnarray}
where $v^{\underline{p-1}}$ can be further divided during this stage: 
$v^{\underline{p-1}}=(v^{\overline{p}}, v^{\underline{p}})$.  
In the end of the Dirac procedure one obtains the following division:
$v^1=(v^{\overline{2}}, v^{\overline{3}}, \ldots ,v^{\overline{N+1}}, 
v^{\underline{N+1}})$, where $v^{\underline{N+1}}$ are 
Lagrangian multipliers remaining undetermined in the process, see next section.

c) With group of $p$-stage Dirac constraints (see next section) we 
assign "contravariant" index $\Phi_{\alpha_p}\equiv \Phi_p$. 
Then complete set of functionally independent constraints of the theory is 
$\Phi_1, \Phi_2, \ldots , \Phi_N$.\par  

d) With a group of functions $\{\psi\}$, first appeared on the stage $p$, we assign 
the symbol $(p)$: $\psi_{(p)}$ (when confusion is not possible, it can be omitted).\par

e) According to these notations, $p$-stage Dirac functions are
\begin{eqnarray}\label{502}
\{\Phi_{\alpha_p}, \Phi_\beta v^\beta +H_0\} \longrightarrow
\{\Phi_{p-1}, \Phi_{1} v^{1} +H_0\}= \cr 
\{\Phi_{p-1}, \Phi_{1}\} v^{1} +\{\Phi_{p-1}, H_0\}\equiv
\triangle_{(p)p-1,1}v^{1}+H_{(p) p-1}.
\end{eqnarray}

f) If $k< p-1$, the notations $\psi_{(p) 1}, \psi_{(p) \underline{k}}$ mean
\begin{eqnarray}\label{503}
\psi_{(p) \underline{k}}=(\psi_{(p) \overline{k+1}}, \psi_{(p) \overline{k+2}}, \ldots ,
\psi_{(p) \overline{p-1}}, \psi_{(p) \underline{p-1}}),
\end{eqnarray}
i.e. the same as for $v^{\underline k}$ in Eq.(\ref{501}).
During the stage $p$, index $\underline{p-1}$ can be further divided 
\begin{eqnarray}\label{504}
\psi_{(p) \underline{p-1}}=(\psi_{(p) \overline{p}}, \psi_{(p) \underline{p}}),
\end{eqnarray}
where (see next section) $\left[\overline{p}\right]$ is number of Lagrangian multipliers determined 
on the stage $p$, and $\left[\underline{p}\right]$ is number of multipliers which remains undetermined 
after the stage $p$.

g) On each stage $p$ it is appear invertible matrix with natural block structure: 
\begin{eqnarray}\label{505}
K_{(p) \widehat{p-1}}{}^{p-1}=
\left(K_{(p) \overline{p}}{}^{p-1}\atop  K_{(p) \widetilde{p}}{}^{p-1}\right)
\end{eqnarray}
It can be used to convert any quantity $\psi_{p-1}$ into $\widehat\psi_{\widehat{p-1}}$ as follow:
\begin{eqnarray}\label{506}
\psi_{p-1}\rightarrow\widehat\psi_{\widehat{p-1}}\equiv K_{(p) \widehat{p-1}}{}^{p-1}\psi_{p-1}=
(\widehat\psi_{\overline{p}}, \widehat\psi_{\widetilde{p}})
\end{eqnarray}
Where confusion is not possible, we write the quantity without hat: 
$\widehat\psi_{\widehat{p-1}}\equiv\psi_{\widehat{p-1}}$. Index $\widetilde{p}$ can be further divided
\begin{eqnarray}\label{507}
\widehat\psi_{\widetilde{p}}=(\widehat\psi_p, \widehat\psi_{\breve p}), 
\Longrightarrow 
\widehat\psi_{\widehat{p-1}}=(\widehat\psi_{\overline p}, \widehat\psi_p, \widehat\psi_{\breve p}),
\end{eqnarray}
where $[p]=[\Phi_p]$, and $\left[\breve p\right]=
\left[ p-1\right]-\left[\overline{p}\right]-\left[ p\right]$.

h) Repeated up and down number of stage imply summation over the corresponding indices. In contrast, 
summation over stages always indicated explicitly, for example
\begin{eqnarray}\label{508}
\sum_{n=2}^p{Q^{(p-n) n}\Phi_n}=
\sum_{n=2}^p \left(\sum_{\alpha_n=1}^{\left[\alpha_n\right]} Q^{(p-n) \alpha_n}\Phi_{\alpha_n}\right).
\end{eqnarray}

In resume, evolution of indices during the stage $p$ can be described as follow:\par
\noindent
For $k< p-1$, the notations $\psi_{(p) 1}, \psi_{(p) \underline{k}}$ are explained in f).\par
\noindent
Index $\underline{p-1}$ of previous stage can be divided on two subgroups 
\begin{eqnarray}\label{509}
\underline{p-1}=(\overline{p}, \underline{p}).
\end{eqnarray}
\noindent
Index ${p-1}$ of previous stage can be converted into $\widehat{p-1}$ and then divided on three 
subgroups 
\begin{eqnarray}\label{510}
p-1\rightarrow\widehat{p-1}=(\overline{p}, p, \breve p).
\end{eqnarray}
As it will be shown in the next section, dimensions of the indices are related with 
rank properties of $p$-stage Dirac system as follows:\par
\noindent
$\left[\overline{p}\right]$ is number of Lagrangian multipliers which are determined on 
the stage $p$;\par
\noindent
$\left[\underline{p}\right]$ is number of multipliers which remains undetermined after 
the stage $p$;\par  
\noindent
$[p]$ is number of $p$-stage Dirac constraints;\par 
\noindent
$\left[\breve p\right]=\left[p-1\right]-\left[\overline{p}\right]-[p]$ is called {\sf{defect}} of 
$p$-stage Dirac system ({\ref{502}). Number of independent (but possibly reducible) $p$-stage 
symmetries, which can be find by our procedure, coincides with the defect $\left[\breve p\right]$, 
see below.

\section{Normal form of $p$-stage Dirac functions}

Primary Dirac constraints have been specified in section 2. Aim of this section is to give formal definition 
for $p$-stage Dirac constrains by induction. The definition is based on possibility to rewrite 
$p$-stage Dirac functions $\{\Phi_{p-1}, H\}$ in special form (\ref{601}) which will be called 
{\tt normal form}. 
The normal form will be our basic expression for analysis of the generating equations below. 

{\bf{Definition.}} 1) First-stage constraints are (functionally independent) primary constraints: 
$\Phi_1\equiv\Phi_{\alpha}$.\par
\noindent
2) Let $\Phi_1, \Phi_2, \ldots \Phi_{p-1}$ is set of constraints of previous stages. Suppose that  
$p$-stage Dirac functions have been {\em{identically}} rewritten in the form 
\begin{eqnarray}\label{601}
\{\Phi_{p-1}, H\}= \qquad \qquad \qquad \qquad \cr \\[5pt] \nonumber
A_{(p) p-1}{}^{\widehat{p-1}} 
\left(
\begin{array}{c}
\pi_{\overline{p}}\\
\Phi_p(q^A, p_j)+B_{(p) p}(\pi_{\overline{p-1}}, \ldots ,\pi_{\overline{2}})\\
C_{(p)\breve p}(\Phi_{p-1}, \ldots ,\Phi_2)+
D_{(p)\breve p}(\pi_{\overline{p-1}}, \ldots ,\pi_{\overline 2})
\end{array}
\right)
\end{eqnarray}
where\par
\noindent
a) Functions $\pi_{\overline k}(v^{\underline{k-1}}), ~ k=2, 3, \ldots ,p$ have the structure
\begin{eqnarray}\label{602}
\pi_{\overline{2}}=X_{(2)\bar 2 1}v^1+Y_{(2) \bar 2}, \qquad \qquad \qquad \qquad \qquad \qquad \
\qquad \cr
\pi_{\overline k}=X_{(k) \overline{k} \underline{k-1}}v^{\underline{k-1}}+
Y_{(k) \overline{k}}, \qquad 
rank\frac{\partial\pi_{\overline k}}{\partial v^{\overline k}}=\left[\pi_{\overline k}\right]=
\left[v^{\overline k}\right], 
\end{eqnarray}
with some coefficients $X(z_1), ~ Y(z_1)$. 
\par
\noindent
b) $B_{(p) p}(\pi_{\overline{p-1}}, \ldots ,\pi_{\overline{2}})$, ~   
$C_{(p)\breve p}(\Phi_{p-1}, \ldots ,\Phi_2)$, ~   
$D_{(p)\breve p}(\pi_{\overline{p-1}}, \ldots ,\pi_{\overline 2})$ 
are linear homogeneous functions of indicated variables, with coefficients dependent 
on $z_1\equiv(q^A, p_j)$ only.\par 
\noindent
c) The matrix $A_{(p) p-1}{}^{\widehat{p-1}}(q^A, p_j)$ is invertible.\par 
\noindent
d) The group $z_1$ is divided on $z_1=(\bar z_2, \bar z_3, \ldots , \bar z_p, z_p)$, such that 
\begin{eqnarray}\label{603}
rank\frac{\partial\Phi_k}{\partial z_1}\Biggr|_{\Phi}=
rank\frac{\partial\Phi_{k}}{\partial \bar z_k}\Biggr|_{\Phi}=
\left[\bar z_k\right]=\left[\Phi_k\right], ~ k=2, 3, \ldots ,p, \cr
rank\frac{\partial (\Phi_2, \ldots , \Phi_p)}{\partial z_1}\Biggr|_{\Phi}=
rank\frac{\partial (\Phi_2, \ldots , \Phi_p)}{\partial (\bar z_2, \ldots \bar z_p})\Biggr|_{\Phi}= \cr 
\left[\bar z_2\right]+ \ldots +\left[\bar z_p\right]=\left[\Phi\right],
\end{eqnarray}
where $\Phi=(\Phi_2, \ldots ,\Phi_p)$. It means that
$\Phi_p(z_1)$ are functionally independent, and $\Phi_k$, $k=(1, 2, \ldots , p)$ are 
functionally independent functions also (note that the primary constraints are included).\par
\noindent
Then the functions $\Phi_p(q^A, p_j)$ are called {\sf{$p$-stage constraints}}. 

The matrix $A$ and 
matrices which form $B, C, D$ will be called {\sf{$p$-stage structure matrices}}.  

To confirm the definition, we use induction over number of stage $p$ to prove that 
the Dirac functions can be actually rewritten in the normal form (\ref{601}).\par
\noindent
1) It was demonstrated in section 3 that second-stage Dirac functions can be presented in the form 
(\ref{601}), see Eqs.(\ref{413}), (\ref{404}). On the case, one has $B=C=D=0$.\par 
\noindent
2) Suppose the Dirac functions of stages $k=2, 3, \ldots ,p-1$ have been presented in the normal form, 
and thus the constraints $\Phi_1, \Phi_2$, $\ldots ,\Phi_{p-1}$ are specified according to 
the definition. Let us consider $p$-stage Dirac functions 
\begin{eqnarray}\label{604}
\{\Phi_{p-1}, \Phi_{1} v^{1} +H_0\}= \qquad \qquad \qquad \qquad \cr 
\{\Phi_{p-1}, \Phi_{1}\} v^{1} +\{\Phi_{p-1}, H_0\}\equiv
\nabla_{(p)p-1,1}v^{1}+H_{(p) p-1}.
\end{eqnarray}
According to the induction hypothesis a), one has the division $v^1=(v^{\overline{2}}, v^{\underline{2}})$,       
$v^{\underline{2}}=(v^{\overline{3}}$, $v^{\underline{3}}), \ldots , 
v^{\underline{p-2}}=(v^{\overline{p-1}}, v^{\underline{p-1}})$, where, using Eq.(\ref{602}), each 
$v^{\overline k}$ can be identically written in the form (see also Lemma1 in Appendix A) 
\begin{eqnarray}\label{605}
v^{\overline{k}}\equiv\widetilde X_{(k)}{}^{\overline{k} \overline{k'}}
\pi_{\overline{k'}}(v^{\underline{k-1}})+
\Lambda_{(k)}{}^{\overline{k}}{}_{\underline{k}}v^{\underline{k}}+W_{(k)}{}^{\overline{k}}.
\end{eqnarray}
Using this representation in Eq.(\ref{604}) one obtains    
\begin{eqnarray}\label{606}
\{\Phi_{p-1}, H\}=\triangle_{(p)p-1 \underline{p-1}}v^{\underline{p-1}}+L_{(p) p-1}+
M_{(p) p-1}(\pi_{\overline{p-1}}, \ldots ,\pi_{\overline 2}), 
\end{eqnarray}
where $\triangle, ~ L, ~ M$ can be find in terms of $\nabla_{(p)}, H_{(p)}$ with help of 
recurrence relations, 
see Lemma 2 in Appendix A. Further, the conversion matrix $K_{(p)}$, 
constructed starting from $\triangle_{(p)}$, can be used to separate $v$-dependent and 
$v$-independent functions among (\ref{606}). According to Lemma 1 of Appendix A one obtains 
\begin{eqnarray}\label{607}
\{\Phi_{p-1}, H\}=
\widetilde K_{(p) p-1}{}^{\widehat{p-1}}\left(\pi_{\overline{p}}(v^{\overline{p-1}})\atop 
\Phi_{\widetilde{p}}(q^A, p_j)+
N_{\widetilde p}(\pi_{\overline{p-1}}, \ldots ,\pi_{\overline 2})\right),
\end{eqnarray}
where
\begin{eqnarray}\label{608}
\begin{array}{l}
\pi_{\overline{p}}=X_{(p) \overline p \underline{p-1}}v^{\underline{p-1}}+Y_{(p) \overline p}, \cr 
\Phi_{\widetilde{p}}\equiv K_{(p) \widetilde p}{}^{p-1} L_{(p) p-1}, \cr 
N_{\widetilde p}=K_{(p) \widetilde p}{}^{p-1}M_{(p) p-1}.
\end{array}
\end{eqnarray}
Here it was denoted 
\begin{eqnarray}\label{609}
\begin{array}{l}
X_{(p) \overline p \underline{p-1}}=K_{(p) \overline p}{}^{p-1}
\triangle_{(p) p-1 \underline{p-1}},\qquad 
rank X_{(p)}=\left[\pi_{\overline p}\right], \cr 
Y_{(p) \overline p}=K_{(p) \overline p}{}^{p-1}L_{(p) p-1}. \qquad \qquad \qquad \qquad
\qquad \qquad \qquad
\end{array}
\end{eqnarray}
Let us analyze the functions $\Phi_{\widetilde{p}}(z_1)$ in Eq.(\ref{607}). Some of them may be 
functionally independent on the constraints of previous stages as well as functionally independent 
among themselves. Induction hypothesis d) allows one to write the representation (see Lemma 3 in 
Appendix A)
\begin{eqnarray}\label{610}
\Phi_{\widetilde{p}}(z_1)=U_{(p) \widetilde{p}}{}^{\widetilde{p}}
\left(\Phi_p(z_1)\atop C_{\breve p}(\Phi_{p-1}, \ldots ,\Phi_2)\right), \cr 
rank\frac{\partial\Phi_p}{\partial\bar z_{p-1}}\Biggr|_{\Phi}=\left[\Phi_p\right],
\qquad \det U\ne 0,
\end{eqnarray}
where index $\widetilde p$ was divided on two groups $\widetilde p=(p, \breve p)$, and the rank condition 
implies that $z_{p-1}$ can be divided: $z_{p-1}=(\bar z_p, z_p)$, 
$\left[\bar z_p\right]=\left[\Phi_p\right]$, $\bar z_p$ can be find through $z_p$ 
from the equations $\Phi_p=0$.  

Substitution of Eq.(\ref{610}) into Eq.(\ref{607}) gives the normal form (\ref{601}) of 
second-stage Dirac functions, with the quantities $A_{(p)}, ~ B_{(p)}, ~ D_{(p)}$ being 
\begin{eqnarray}\label{611}
A_{(p) p-1}{}^{\widehat{p-1}}=\widetilde K_{(p)}
\left(
\begin{array}{cc}
1_{(\overline{p})\times(\overline{p})}&0\\
0&U_{(p) \widetilde{p}}{}^{\widetilde{p}}
\end{array}
\right),
\end{eqnarray}
\begin{eqnarray}\label{612}
B_{(p) p}=(\widetilde U_{(p)}N)_p, \quad 
D_{(p) \breve p}=(\widetilde U_{(p)}N)_{\breve p}.
\end{eqnarray}
It finishes the proof. We have described procedure to represent $p$-stage Dirac functions in 
the normal form (\ref{601}). 

It is easy to see that our definition of $p$-stage constraints is equivalent to the 
standard one. Actually, according to Dirac, to reveal $p$-stage constraints one studies the equations 
$\{\Phi_{p-1}, H\}$$=0$ on the surface of previously determined constraints and Lagrangian multipliers. 
But, according to our proof, on this surface the equations are equivalent to the system 
$\pi_{\overline p}(v)=0, ~ \Phi_p(q^A, p_j)=0$, where the first equation determines some of Lagrangian 
multipliers while the second one represents our $p$-stage constraints. Note that division 
on subgroups has been made in accordance with rank properties of $p$-stage Dirac system (\ref{601}), 
which determines dimensions of subgroups as they were described in the end of section 5.  

\section{Normal form of second and third stage generating functions}

In the next section we develop procedure to rewrite the set of $s$-stage generating functions 
(\ref{303}) in the {\sf normal form} (\ref{802}), i.e. as combination of constraints. Before doing this, it is 
instructive to see how  the procedure works for lower stages. 

Let us consider second-stage generating function
\begin{eqnarray}\label{701}
T^{(2)}=Q^{(2)1}\{\Phi_1, H\},
\end{eqnarray}
where the coefficients $Q^{(2) 1}(q^A, p_j)$ are arbitrary functions. Using Eq.(\ref{413}) 
one obtains  
\begin{eqnarray}\label{702}
T^{(2)}=\widehat Q^{(2) \overline 2}\pi_{\overline 2}+
\widehat Q^{(2) 2}\Phi_2+\widehat Q^{(2) \breve 2}0_{\breve 2}, 
\end{eqnarray}
where $\widehat Q$ represent bloks of converted $Q$:
\begin{eqnarray}\label{703}
Q^{(2)1}A_{(2) 1}{}^{\widehat 1}\equiv \widehat Q^{(2) \widehat 1}=(\widehat Q^{(2) \overline 2}, ~  
\widehat Q^{(2) 2}, ~ \widehat Q^{(2) \breve 2}).  
\end{eqnarray}
Let us take $\widehat Q^{(2) \overline 2}=0$ and denote $\widehat Q^{(2) 2}=-Q^{(2) 2}$. It leads to 
the desired result
\begin{eqnarray}\label{704}
T^{(2)}=-Q^{(2) 2}\Phi_2.
\end{eqnarray}
It was achieved by the following choice:
\begin{eqnarray}\label{705}
Q^{(2)1}=(0^{\overline 2}, ~ -Q^{(2) 2}, ~ \widehat Q^{(2) \breve 2})\widetilde A_{(2) \widehat 1}{}^1,
\end{eqnarray}
where $Q^{(2) 2}, ~ \widehat Q^{(2) \breve 2}$ remains arbitrary functions. 

Taking further 
$Q^{(2) 2}=0$, one obtains solution (\ref{705}) of second-stage generating equations (\ref{309}). 
According to Statement 1, one writes out immediately $\left[\breve 2\right]$ independent local 
symmetries of the Lagrangian action, see Sect. 9. Number of second-stage symmetries coincides with 
the defect of second-stage Dirac system (\ref{413}). 
Note also that the symmetries are specified, in fact, by second-stage structure matrix $A_{(2)}$.

Let us consider set of third-stage generating functions
\begin{eqnarray}\label{706}
T^{(2)}=Q^{(2)1}\{\Phi_1, H\},
\end{eqnarray}
\begin{eqnarray}\label{707}
T^{(3)}=Q^{(3) 1}\{\Phi_1, H\}+\{H, T^{(2)}\},
\end{eqnarray}
where the coefficients $Q^{(2) 1}(q^A, p_j, v^{\alpha}), ~ Q^{(3) 1}(q^A, p_j, v^{\alpha})$ are 
some functions.   
As in the previous case, making the choice (\ref{705}), one writes $T^{(2)}$ in the normal 
form (\ref{704}). Using this expression for $T^{(2)}$ as well as Eq.(\ref{601}), one obtains the following 
expression for $T^{(3)}$ 
\begin{eqnarray}\label{708}
T^{(3)}=Q^{(3) 1}\{\Phi_1, H\}-\{H, Q^{(2) 2}\Phi_2\}= \cr 
\widehat Q^{(3) \widehat 1}
\left(
\begin{array}{c}
\pi_{\overline{2}}\\
\Phi_2\\
0_{\breve 2}
\end{array}
\right)-\{H, Q^{(2)2}\}\Phi_2+ \\[3pt] \nonumber
\widehat Q^{(2) \widehat 2}
\left(
\begin{array}{c}
\pi_{\overline{3}}\\
\Phi_3+B_{(3) 3}(\pi_{\overline{2}})\\
C_{(3)\breve 3}(\Phi_2)+
D_{(3)\breve 3}(\pi_{\overline 2})
\end{array}
\right) 
\end{eqnarray}
where 
\begin{eqnarray}\label{709}
Q^{(2) 2}A_{(3) 2}{}^{\widehat 2}\equiv\widehat Q^{(2) \widehat 2}=(\widehat Q^{(2) \overline 3}, ~  
\widehat Q^{(2) 3}, ~ \widehat Q^{(2) \breve 3}), \\[1pt] \nonumber   
Q^{(3)1}A_{(2) 1}{}^{\widehat 1}\equiv\widehat Q^{(3) \widehat 1}=(\widehat Q^{(3) \overline 2}, ~  
\widehat Q^{(3) 2}, ~ \widehat Q^{(3) \breve 2}).  
\end{eqnarray}
Collecting similar terms in Eq.(\ref{708}) one has
\begin{eqnarray}\label{710}
T^{(3)}=
\left(\widehat Q^{(3) \overline 2}+\widehat Q^{(2) 3}B_{(3) 3}{}^{\overline 2}+
\widehat Q^{(2) \breve 3}D_{(3)\breve 3}{}^{\overline 2}\right)\pi_{\overline{2}}+
\widehat Q^{(2) \overline 3}\pi_{\overline{3}}+ \\[1pt] \nonumber
\left(\widehat Q^{(3) 2}-\{H, Q^{(2)2}\}+\widehat Q^{(2) \breve 3}C_{(3)\breve 3}{}^2\right)\Phi_2+
\widehat Q^{(2) 3}\Phi_3+
\widehat Q^{(3) \breve 2}0_{\breve 2}.
\end{eqnarray}
Then the following choice 
\begin{eqnarray}\label{711}
\widehat Q^{(2) \overline 3}=0, ~  
\widehat Q^{(3) \overline 2}=-\widehat Q^{(2) 3}B_{(3) 3}{}^{\overline 2}-
\widehat Q^{(2) \breve 3}D_{(3)\breve 3}{}^{\overline 2}, \qquad \quad \\[1pt] \nonumber
\widehat Q^{(2) 3}\equiv -Q^{(2) 3}, ~  
\widehat Q^{(3) 2}=-Q^{(3) 2}+\{H, \widehat Q^{(2) \widehat 2}\widetilde A_{(3) \widehat 2}{}^2\}-
\widehat Q^{(2) \breve 3}C_{(3)\breve 3}{}^2,
\end{eqnarray}
with arbitrary functions $Q^{(2) 3}, ~ Q^{(3) 2}$, gives $T^{(3)}$ in the normal form  
\begin{eqnarray}\label{712}
T^{(3)}=-Q^{(3) 2}\Phi_2-Q^{(2) 3}\Phi_3
\end{eqnarray}
Thus the normal form (\ref{704}), (\ref{712}) for the third-stage generating functions is supplied by  
special choice of $Q^{(2) 1}, ~ Q^{(3) 1}$. The coefficients have been divided on the following groups: 
\begin{eqnarray}\label{7133}
Q^{(2) 1}=\left(0^{\overline 2}, \left(0^{\overline 3}, \widehat Q^{(2) 3}, 
\widehat Q^{(2) \breve 3}\right)\widetilde A_{(3) \widehat 2}{}^2, \widehat Q^{(2) \breve 2}\right)
\widetilde A_{(2) \widehat 1}{}^1, \cr
Q^{(3) 1}=\left(\widehat Q^{(3) \overline 2}, \widehat Q^{(3) 2}, \widehat Q^{(3) \breve 2}\right)
\widetilde A_{(2) \widehat 1}{}^1, \qquad \qquad
\end{eqnarray}
To describe structure of the groups, it will be convenient to use the following triangle table 
\begin{eqnarray}\label{713}
\begin{array}{ccccccccccc}
Q^{(2) 1}&\sim&&&&0^{\overline 2}&0^{\overline 3}&Q^{(2) 3}&\widehat Q^{(2) \breve 3}&
\widehat Q^{(2) \breve 2}\\
Q^{(3) 1}&\sim&&&&&\widehat Q^{(3) \overline 2}&Q^{(3) 2}&\widehat Q^{(3) \breve 2}&\\
\end{array}
\end{eqnarray}
Writting out such a kind  
tables below, we omite the conversion matrices and write arbitrary coefficients $Q^k$ instead of 
$\widehat Q^k$ in central column of the table.
Then the central column and columns on r.h.s. of it represent coefficients which remains arbitrary. 
Detailed expression for 
division of $Q^{(p) 1}$ on subgroups of $s$-stage can be find in Appendix B, see Eq.(\ref{3.10}).

Taking  
$Q^{(2) 3}=0, ~ Q^{(3) 2}=0$, one obtains solution (\ref{713}), (\ref{711}) of third-stage generating 
equations (\ref{310}). 
According to Statement 1, it implies $\left[\breve 3\right]$ independent third-stage local 
symmetries of the Lagrangian action, see Sect. 9. Number of the symmetries coincides with the defect 
of third-stage Dirac system (\ref{601}). 
The symmetries are specified, in fact, by the structure matrices 
$A_{(2)}, A_{(3)}, B_{(3)}, C_{(3)}, D_{(3)}$. 

\section{Normal form of $s$-stage generating functions}

We demonstrate here that the set of $s$-stage generating functions can be written in 
the {\sf normal form} (\ref{802}).

{\bf Statement 3.} Consider the Lagrangian theory with the Hamiltonian $H$, and with constraints 
at most $N$ stage appeared in the Hamiltonian 
formulation: $\Phi_1, \Phi_2, \ldots ,\Phi_N$. For some fixed integer number $s$, $2\le s\le N+1$, let 
us construct the generating 
functions according to recurrence relations ($T^{(1)}=0$)
\begin{eqnarray}\label{801}
T^{(p)}=Q^{(p) 1}\{\Phi_1, H\}+\{H, T^{(p-1)}\}, \qquad p=2,3, \ldots ,s.
\end{eqnarray}
Then the coefficients $Q^{(p)1}(q^A, p_j, v^{\alpha})$ can be chosen in such a way, that 
all $T^{(p)}$ turn out to be linear combinations of the constraints  
\begin{eqnarray}\label{802}
T^{(p)}=-\sum_{n=2}^p Q^{(p+2-n) n}\Phi_n, \qquad p=2,3, \ldots ,s. 
\end{eqnarray}
Choice of $Q^{(p)1}$, which supplies the normal form can be described as follows: \par
\noindent
a) For any $k=2,3, \ldots ,s, ~ n=1,2, \ldots ,(s+1-k)$, $Q^{(k) n}$ is divided on three subgroups  
with help of the structure matrix of $n+1$ stage $A_{(n+1)}$ 
\begin{eqnarray}\label{8033}
Q^{(k) n}A_{(n+1) n}{}^{\widehat n}\equiv\widehat Q^{(k) \widehat n}=
(\widehat Q^{(k) \overline{n+1}}, \widehat Q^{(k) n+1}, \widehat Q^{(k) \breve{n+1}}), 
\end{eqnarray}
where for any $p=2,3, \ldots ,s, ~ n=2,3, \ldots ,p$ one has 
\begin{eqnarray}\label{803}
\widehat Q^{(p+2-n) n}=-Q^{(p+2-n) n}+\{H, \widehat Q^{(p+1-n) \widehat n}
\widetilde A_{(n+1) \widehat n}{}^n\}- \qquad \cr 
\sum_{m=n}^{p-1} \widehat Q^{(p+1-m) \breve{m+1}}C_{(m+1) \breve{m+1}}{}^n, \qquad 
\Longrightarrow \widehat Q^{(2) p}=-Q^{(2) p},
\end{eqnarray}
\begin{eqnarray}\label{804}
\widehat Q^{(p+2-n) \overline n}=-
\sum_{m=n}^{p-1}\left(\widehat Q^{(p+1-m) m+1}B_{(m+1) m+1}{}^{\overline n}+\right. \cr
\left. \widehat Q^{(p+1-m) \breve{m+1}}D_{(m+1) \breve{m+1}}{}^{\overline n}\right) \qquad \quad 
\Longrightarrow \widehat Q^{(2) \overline p}=0,
\end{eqnarray}
and $B, ~ C, ~ D$ are structure matrix of the Dirac procedure, see Eq.(\ref{601}). \par
\noindent
b) The coefficients $\widehat Q^{(k) \breve{n+1}}, ~ k=2,3, \ldots ,s, ~ n=1,2, \ldots ,(s+1-k)$ remain 
arbitrary. \par
\noindent
c) The coefficients $Q^{(s+2-n) n}, ~ n=2,3, \ldots ,s$ remain arbitrary.
  
Before carrying out a proof, let us confirm that the recurrence relations (\ref{803}), (\ref{804}) 
actually determines the coefficients.
According to the statement, $Q^{(p)1}$ is converted into $\widehat Q^{(p) \widehat 1}$, and then is 
divided on subgroups 
$(\widehat Q^{(p) \overline 2}, \widehat Q^{(p) 2}, \widehat Q^{(p) \breve 2})$. Then  
$Q^{(p) 2}$ is picked out from $\widehat Q^{(p) 2}$ according to Eq.(\ref{803}). The coefficient 
$Q^{(p) 2}$ will be further converted and divided, creating $Q^{(p) 3}$, and so on.   
Resulting structure of the coefficients $Q^{(p)1}$ can be described by the following table (first line 
represents $Q^{(2)1}$, second line represents $Q^{(3)1}$, and so on, up to $Q^{(s)1}$): 
\begin{eqnarray}\label{805}
\begin{array}{ccccccccccc}
0^{\overline 2}&0^{\overline 3}&0^{\overline 4}&\ldots&0^{\overline s}&
Q^{(2) s}&\widehat Q^{(2) \breve s}&\ldots&\widehat Q^{(2) \breve 4}&\widehat Q^{(2) \breve 3}&
\widehat Q^{(2) \breve 2}\\
&\widehat Q^{(3)\overline 2}&\widehat Q^{(3)\overline 3}&\ldots&\widehat Q^{(3) \overline{s-1}}&
Q^{(3) s-1}&\widehat Q^{(3) \breve{s-1}}&\ldots&\widehat Q^{(3) \breve 3}&\widehat Q^{(3) \breve 2}&\\
&&\ldots&\ldots&\ldots&\ldots&\ldots&\ldots&\ldots&&\\
&&\widehat Q^{(p)\overline 2}&\ldots&\widehat Q^{(p)\overline{s+2-p}}&Q^{(p) s+2-p}&
\widehat Q^{(p) \breve{s+2-p}}&\ldots&\widehat Q^{(p) \breve 2}&&\\
&&&\ldots&\ldots&\ldots&\ldots&\ldots&&\\
&&&\widehat Q^{(s-1)\overline 2}&\widehat Q^{(s-1)\overline 3}&Q^{(s-1) 3}&
\widehat Q^{(s-1) \breve 3}&\widehat Q^{(s-1) \breve 2}&&&\\
&&&&\widehat Q^{(s)\overline 2}&Q^{(s) 2}&\widehat Q^{(s) \breve 2}&&&&\\
\end{array} 
\end{eqnarray}
Note that any group $Q^{(p)n}$ with $n\ne s+2-p$ is presented on the table by the interval of $p$-line between 
$\widehat Q^{(p) \overline{n+1}}$ and $\widehat Q^{(p) \breve{n+1}}$. Manifest form for 
division of $Q^{(p) 1}$ on subgroups of $s$-stage can be find in Appendix B, see Eq.(\ref{3.10}).

To analyse the expressions (\ref{803}), (\ref{804}) it is convenient to fix number of line: $p+2-n=k$, then they 
can be written in the form
\begin{eqnarray}\label{806}
\widehat Q^{(k) n}=-Q^{(k) n}+\{H, \widehat Q^{(k-1) \widehat n}
\widetilde A_{(n+1) \widehat n}{}^n\}- \qquad \cr 
\sum_{m=2}^{k-1} \widehat Q^{(m) \breve{k+n-m}}C_{(k+n-m) \breve{k+n-m}}{}^n, 
\end{eqnarray}
\begin{eqnarray}\label{807}
\widehat Q^{(k) \overline n}=-
\sum_{m=2}^{k-1}\left(\widehat Q^{(m) k+n-m}B_{(k+n-m) k+n-m}{}^{\overline n}+\right. \cr
\left. \widehat Q^{(m) \breve{k+n-m}}D_{(k+n-m) \breve{k+n-m}}{}^{\overline n}\right),  
\end{eqnarray}
where $k=2,3, \ldots ,s, ~ n=2,3, \ldots ,(s+2-k)$. From this it follows that any group 
$\widehat Q^{(k) \overline n}$ of the line $k$ of the triangle is expressed through some groups placed in 
previous lines on r.h.s. of $\widehat Q^{(k) \overline n}$.  
Any group $\widehat Q^{(k) n}$ is presented through the interval 
$\left[\widehat Q^{(k) \overline{n+1}}, ~ \widehat Q^{(k) \breve{n+1}}\right]$ of the line $k$ 
as well as through some groups of previous lines placed on r.h.s. of $n$ column. After all, all the 
coefficients are expressed through $\widehat Q^{(k) \breve{n}}$, $Q^{(s+2-n) n}\equiv Q^{(n) s+2-n}$,  
which remains arbitrary functions (the latters are placed in the central column and on r.h.s. of it in 
the triangle). 
Note that all arbitrary functions $Q^{(s+2-n) n}$ 
appear in the expression for higher generating function $T^{(s)}$. 

{\bf Proof.} The statement 3 will be demonstrated by induction over $s$, $2\le s\le N+1$. It was shown in 
section 7 (see Eqs.(\ref{704}), (\ref{705})) that the statement is true for $s=2$. Supposing that the 
statement is true for $s=p-1$, let us consider generating functions of the stage $s=p$: 
$T^{(2)}, T^{(3)}, \ldots ,T^{(p-1)},T^{(p)}$. According to induction hypothesis,  
$T^{(2)}, T^{(3)}, \ldots ,T^{(p-1)}$ can be written in the normal form       
\begin{eqnarray}\label{808}
T^{(k)}=-\sum_{n=2}^k Q^{(k+2-n) n}\Phi_n, \qquad k=2,3, \ldots ,p-1. 
\end{eqnarray}
where the coefficients $Q^{(k) 1}, k=2,3. \ldots ,p$ have the structure 
\begin{eqnarray}\label{809}
\begin{array}{ccccccccc}
0^{\overline 2}&0^{\overline 3}&\ldots&0^{\overline{p-1}}&
Q^{(2) p-1}&\widehat Q^{(2) \breve{p-1}}&\ldots&\widehat Q^{(2) \breve 3}&\widehat Q^{(2) \breve 2}\\
&\widehat Q^{(3)\overline 2}&\ldots&\widehat Q^{(3) \overline{p-2}}&
Q^{(3) p-2}&\widehat Q^{(3) \breve{p-2}}&\ldots&\widehat Q^{(3) \breve 2}&\\
&&\ldots&\ldots&\ldots&\ldots&\ldots&&\\
&&&\widehat Q^{(p-1)\overline 2}&Q^{(p-1) 2}&\widehat Q^{(p-1) \breve 2}&&\\
&&&&Q^{(p) 1}&&&&\\
\end{array} 
\end{eqnarray}
with arbitrary functions placed in the central column and on r.h.s. of it. Let us consider the remaining 
generating function 
\begin{eqnarray}\label{810}
T^{(p)}=Q^{(p) 1}\{\Phi_1, H\}-\{H, \sum_{n=2}^{p-1} Q^{(p+1-n) n}\Phi_n\},
\end{eqnarray}
Note that only $Q$ of the central column are presented in this expression. Being arbitrary
functions, these $Q$ can be used to write the expression in the normal form, as it is   
demonstrated in Lemma 4 of Appendix B. According to the Lemma, each coefficient of the central column in 
Eq.(\ref{809}) is divided on three subgroups (\ref{8033}). From comparison of Eqs.(\ref{801}), (\ref{802}) 
with correspoding equations (\ref{3.1}) ,(\ref{3.2}) of Lemma 4 one obtains rules to transform 
quantities of Lemma 4 into our case
\begin{eqnarray}\label{811} 
Q^n\rightarrow Q^{(p+1-n) n}, \Longrightarrow 
\widehat Q^{\widehat n}\rightarrow \widehat Q^{(p+1-n) \widehat n}, \qquad  
Q'^n\rightarrow Q^{(p+2-n) n}. 
\end{eqnarray}
It implies that 
$\widehat Q^{(p+2-n) n}, ~ \widehat Q^{(p+2-n) \overline{n}}$ are determined by 
Eqs.(\ref{803}), (\ref{804}). Note that division (\ref{8033}) has been made in accordance with rank 
properties of $p$-stage Dirac system (\ref{601}), the latter determines dimensions of subgroups 
(see description of the dimensions after Eq.(\ref{510})). $\spadesuit$ 

For completeness, let us present normal form of lower-stage generating functions. \par
\noindent
{\bf{Normal form of second-stage generating function.}} For $s=2$ one obtains immediately (see also 
section 7)
\begin{eqnarray}\label{812}
T^{(2)}=Q^{(2) 1}\{\Phi_1, H\}=-Q^{(2) 2}\Phi_2, 
\end{eqnarray}
where 
\begin{eqnarray}\label{813}
Q^{(2)1}=(0^{\overline 2}, ~ -Q^{(2) 2}, ~ \widehat Q^{(2) \breve 2})\widetilde A_{(2) \widehat 1}{}^1,
\end{eqnarray}
with arbitrary functions $Q^{(2) 2}(q^A, p_j), ~ \widehat Q^{(2) \breve 2}(q^A, p_j)$. \par

\noindent
{\bf{Normal form of third-stage generating functions.}} 
\begin{eqnarray}\label{814}
\begin{array}{l}
T^{(2)}=Q^{(2)1}\{\Phi_1, H\}=\qquad \qquad -Q^{(2) 2}\Phi_2, \cr
T^{(3)}=Q^{(3)1}\{\Phi_1, H\}+\{H, T^{(2)}\}= \qquad \qquad \qquad \qquad \cr
Q^{(3)1}\{\Phi_1, H\}-\{H, Q^{(2) 2}\Phi_2\}= 
-Q^{(3) 2}\Phi_2-Q^{(2) 3}\Phi_3.
\end{array}
\end{eqnarray}
Division of the initial coefficients can be described by the triangle
\begin{eqnarray}\label{815}
\begin{array}{ccccccccccc}
Q^{(2) 1}&\sim&&&&0^{\overline 2}&0^{\overline 3}&Q^{(2) 3}&\widehat Q^{(2) \breve 3}&
\widehat Q^{(2) \breve 2}\\
Q^{(3) 1}&\sim&&&&&\widehat Q^{(3) \overline 2}&Q^{(3) 2}&\widehat Q^{(3) \breve 2}&\\
\end{array}
\end{eqnarray}
where the central column and columns on r.h.s. of it represent functions which remain arbitrary.
Manifest form of the coefficients which provides the normal form (\ref{814}) is as follows:
\begin{eqnarray}\label{816}
Q^{(2) 1}=\left(0^{\overline 2}, \left(0^{\overline 3}, \widehat Q^{(2) 3}, 
\widehat Q^{(2) \breve 3}\right)\widetilde A_{(3) \widehat 2}{}^2, \widehat Q^{(2) \breve 2}\right)
\widetilde A_{(2) \widehat 1}{}^1, \cr
Q^{(3) 1}=\left(\widehat Q^{(3) \overline 2}, \widehat Q^{(3) 2}, \widehat Q^{(3) \breve 2}\right)
\widetilde A_{(2) \widehat 1}{}^1, \qquad \qquad
\end{eqnarray}
\begin{eqnarray}\label{817}
\widehat Q^{(2) 3}= -Q^{(2) 3}, \qquad \qquad \quad \qquad \quad \qquad \quad \cr
\widehat Q^{(3) \overline 2}=Q^{(2) 3}B_{(3) 3}{}^{\overline 2}-
\widehat Q^{(2) \breve 3}D_{(3)\breve 3}{}^{\overline 2}, \qquad \qquad \qquad \cr
\widehat Q^{(3) 2}=-Q^{(3) 2}+
\{H, \left(0^{\overline 3}, -Q^{(2) 3}, \widehat Q^{(2) \breve 3}\right)
\widetilde A_{(3) \widehat 2}{}^2\}-
\widehat Q^{(2) \breve 3}C_{(3)\breve 3}{}^2. 
\end{eqnarray}
Here all $Q$ on r.h.s. of Eq.(\ref{817}) are arbitrary functions.
 
The coefficients appeared in $T^{(3)}$ remains arbitrary functions, while $Q^{(2) 2}$ in $T^{(2)}$ is
\begin{eqnarray}\label{818}
Q^{(2) 2}=\left(0^{\overline 3}, -Q^{(2) 3}, 
\widehat Q^{(2) \breve 3}\right)\widetilde A_{(3) \widehat 2}{}^2. 
\end{eqnarray}

\noindent
{\bf{Normal form of 4-stage generating functions.}}
\begin{eqnarray}\label{819}
\begin{array}{l}
T^{(2)}=Q^{(2)1}\{\Phi_1, H\}= \qquad \qquad \quad -Q^{(2) 2}\Phi_2, \qquad \cr
T^{(3)}=Q^{(3)1}\{\Phi_1, H\}+\{H, T^{(2)}\}= \qquad \qquad \qquad \qquad \qquad \cr
Q^{(3)1}\{\Phi_1, H\}-\{H, Q^{(2) 2}\Phi_2\}=
-Q^{(3) 2}\Phi_2-Q^{(2) 3}\Phi_3, \cr
T^{(4)}=Q^{(4)1}\{\Phi_1, H\}+\{H, T^{(3)}\}=
Q^{(4)1}\{\Phi_1, H\}- \cr 
\{H, Q^{(3) 2}\Phi_2+Q^{(2) 3}\Phi_3\}=
-Q^{(4) 2}\Phi_2-Q^{(3) 3}\Phi_3-Q^{(2) 4}\Phi_4.
\end{array}
\end{eqnarray}
Division of the initial coefficients can be described by the triangle
\begin{eqnarray}\label{820}
\begin{array}{ccccccccccccc}
Q^{(2) 1}&\sim&&&&0^{\overline 2}&0^{\overline 3}&0^{\overline 4}&Q^{(2) 4}&\widehat Q^{(2) \breve 4}
&\widehat Q^{(2) \breve 3}&\widehat Q^{(2) \breve 2}\\
Q^{(3) 1}&\sim&&&&&\widehat Q^{(3) \overline 2}&\widehat Q^{(3) \overline 3}&Q^{(3) 3}&\widehat Q^{(3) \breve 3}
&\widehat Q^{(3) \breve 2}&\\
Q^{(4) 1}&\sim&&&&&&\widehat Q^{(4) \overline 2}&Q^{(4) 2}&\widehat Q^{(4) \breve 2}&&\\
\end{array}
\end{eqnarray}
where the central column and columns on r.h.s. of it are arbitrary functions.
Manifest form of the coefficients which provides the normal form (\ref{819}) is as follows:
\begin{eqnarray}\label{821}
Q^{(2) 1}=\left(0^{\overline 2}, \left(0^{\overline 3}, \left(0^{\overline 4}, \widehat Q^{(2) 4},
\widehat Q^{(2) \breve 4}\right)\widetilde A_{(4) \widehat 3}{}^3, 
\widehat Q^{(2) \breve 3}\right)\widetilde A_{(3) \widehat 2}{}^2,
\widehat Q^{(2) \breve 2}\right)\widetilde A_{(2) \widehat 1}{}^1, \cr
Q^{(3) 1}=\left(\widehat Q^{(3) \overline 2}, \left(\widehat Q^{(3) \overline 3}, \widehat Q^{(3) 3},
\widehat Q^{(3) \breve 3}\right)\widetilde A_{(3) \widehat 2}{}^2,
\widehat Q^{(3) \breve 2}\right)\widetilde A_{(2) \widehat 1}{}^1, \qquad \qquad \cr
Q^{(4) 1}=\left(\widehat Q^{(4) \overline 2}, \widehat Q^{(4) 2}, \widehat Q^{(4) \breve 2}\right)
\widetilde A_{(2) \widehat 1}{}^1. \qquad \qquad \qquad \qquad \qquad 
\end{eqnarray}
\begin{eqnarray}\label{822}
\begin{array}{l}
\widehat Q^{(2) 4}=-Q^{(2) 4}, \qquad \qquad \qquad \qquad \qquad \qquad  
\qquad \qquad \qquad \qquad \qquad \cr 
\widehat Q^{(3) \overline 2}=\left(0^{\overline 4}, -Q^{(2) 4}, 
\widehat Q^{(2) \breve 4}\right)\widetilde A_{(4) \widehat 3}{}^3       
B_{(3) 3}{}^{\overline 2}-
\widehat Q^{(2) \breve 3}D_{(3)\breve 3}{}^{\overline 2}, \qquad \qquad \qquad \cr 
\widehat Q^{(3) \overline 3}=Q^{(2) 4}B_{(4) 4}{}^{\overline 3}-
\widehat Q^{(2) \breve 4}D_{(4)\breve 4}{}^{\overline 3}, 
\qquad \qquad \qquad \qquad \qquad \qquad \qquad \cr
\widehat Q^{(3) 3}=-Q^{(3) 3}+\{H, \left(0^{\overline 4}, -Q^{(2) 4}, 
\widehat Q^{(2) \breve 4}\right)\widetilde A_{(4) \widehat 3}{}^3\}-
\widehat Q^{(2) \breve 4}C_{(4) \breve 4}{}^3, \qquad \cr
\widehat Q^{(4) \overline 2}=Q^{(2) 4}B_{(4) 4}{}^{\overline 2}-
\widehat Q^{(2) \breve 4}D_{(4)\breve 4}{}^{\overline 2}- 
\left(-Q^{(3) 3}+ \right. \qquad \qquad \qquad \qquad \quad \cr 
\left.\{H, \left(0^{\overline 4}, -Q^{(2) 4}, 
\widehat Q^{(2) \breve 4}\right)\widetilde A_{(4) \widehat 3}{}^3\}-
\widehat Q^{(2) \breve 4}C_{(4) \breve 4}{}^3\right)B_{(3) 3}{}^{\overline 2}-
\widehat Q^{(3) \breve 3}D_{(3)\breve 3}{}^{\overline 2}, \cr  
\widehat Q^{(4) 2}=-Q^{(4) 2}+\{H, \left(
Q^{(2) 4}B_{(4) 4}{}^{\overline 3}-\widehat Q^{(2) \breve 4}D_{(4)\breve 4}{}^{\overline 3}, \right. \cr 
\left.-Q^{(3) 3}+\{H, \left(0^{\overline 4}, -Q^{(2) 4}, 
\widehat Q^{(2) \breve 4}\right)\widetilde A_{(4) \widehat 3}{}^3\}-
\widehat Q^{(2) \breve 4}C_{(4) \breve 4}{}^3, \right. \cr  
\left.\widehat Q^{(3) \breve 3}\right)\widetilde A_{(3) \widehat 2}{}^2\}-
\widehat Q^{(2) \breve 4}C_{(4) \breve 4}{}^2-\widehat Q^{(3) \breve 3}C_{(3) \breve 3}{}^2.
\end{array}
\end{eqnarray}
Here all $Q$ on r.h.s. of Eq.(\ref{822}) are arbitrary functions.

The coefficients $Q$ in $T^{(4)}$ remain arbitrary functions.
The line between $0^{\overline 3}$ and $\widehat Q^{(2) \breve 3}$ in $Q^{(2) 1}$ represents 
the coefficient $Q^{(2) 2}$ which appears in $T^{(2)}$, similarly can be find all the coefficients 
appeared in $T^{(3)}$. One obtains their manifest form as follows:
\begin{eqnarray}\label{823}
\begin{array}{l}
Q^{(2) 2}=\left(0^{\overline 3}, \left(0^{\overline 4}, -Q^{(2) 4},
\widehat Q^{(2) \breve 4}\right)\widetilde A_{(4) \widehat 3}{}^3, 
\widehat Q^{(2) \breve 3}\right)\widetilde A_{(3) \widehat 2}{}^2, \qquad \qquad \quad \cr
Q^{(2) 3}=\left(0^{\overline 4}, -Q^{(2) 4},
\widehat Q^{(2) \breve 4}\right)\widetilde A_{(4) \widehat 3}{}^3, 
\qquad \qquad \qquad \qquad \qquad \qquad \quad\cr
Q^{(3) 2}=\left(Q^{(2) 4}B_{(4) 4}{}^{\overline 3}-
\widehat Q^{(2) \breve 4}D_{(4)\breve 4}{}^{\overline 3}, -Q^{(3) 3}+ \right.  \qquad \qquad \qquad\cr 
\left. \{H, \left(0^{\overline 4}, -Q^{(2) 4}, 
\widehat Q^{(2) \breve 4}\right)\widetilde A_{(4) \widehat 3}{}^3\}-
\widehat Q^{(2) \breve 4}C_{(4) \breve 4}{}^3,
\widehat Q^{(3) \breve 3}\right)\widetilde A_{(3) \widehat 2}{}^2.
\end{array}
\end{eqnarray}

\section{Gauge symmetries of quadratic theory}

Suppose that in the Hamiltonian formulation of our theory there are appear constraints up to at 
most $N$-stage. According to the Statement 1, symmetries of different stages are looked for separately. 
Generators of $s$-stage local symmetries (\ref{306})-(\ref{308}) can be constructed starting from any 
solution of generating equations (\ref{304})-(\ref{305}). Using normal form (\ref{802})-(\ref{807}) of 
generating functions, 
one concludes that Eq.(\ref{305}) is satisfied by taking $Q^{(p) s+2-p}=0, ~ p=2, 3, \ldots ,s$, i.e. all 
the coefficients in the central column of the triangle (\ref{805}) must be zeros. 
Eq.(\ref{304}) states that generating functions $T^{(p)}$ with $p=2, 3, \ldots ,s-1$ do not depend on 
the Lagrangian multipliers. Dependence on $v^1$ can appear only due to second term in Eq.(\ref{803})
(or, equivalently, in Eq.(\ref{806})). Thus one needs to kill this term, which can be easily achieved in 
a theory with all the structure matrices $\widetilde A$ (see Eq.(\ref{601})) being numerical matrices.  
It happens, in particular, in any quadratic theory (then all the structure matrices 
$A, B, C, D$ in Eq.(\ref{601}) turn out to be numerical matrices). 
We analyse this case in the present section. For the case, it is consistent to look for solutions with 
$Q=const$, then the second term in Eq.(\ref{803}) disappears, and the generating equations 
(\ref{304})-(\ref{305}) are trivially satisfied. 

Thus for any quadratic Lagrangian theory it is sufficient to take elements of central column of 
the triangle (\ref{805}) be zeros, and elements on r.h.s. of it be arbitrary numbers, to obtain some 
local symmetry of the Lagrangian action (\ref{307}), (\ref{308}). In the case, the generators turn out 
to be numerical matrices. 

Let us discuss some particular set of generators constructed as follows. On the stage $s$ of the 
Dirac procedure, one takes $\left[\breve {s'}\right]$ sets of  $\widehat Q^{(2)\breve s}$, namely 
$\widehat Q^{(2)}{}_{\breve {s'}}{}^{\breve s}=\delta_{\breve {s'}}{}^s$, where 
$\left[\breve{s'}\right]$ is defect of 
the system (\ref{601}). Remaining arbitrary coefficients on r.h.s. of the triangle (\ref{805}) 
are taken vanishing. Then these $\left[\breve{s'}\right]$ solutions 
$Q^{(p)}{}_{\breve{s'}}{}^{1}, p=2, 3, \ldots ,s$ of generating equations have the form (\ref{3.10}) of 
Appendix C, where one needs to substitute 
\begin{eqnarray}\label{901}
Q^{(p) s+2-p}=0, \qquad \widehat Q^{(2)}{}_{\breve{s'}}{}^{\breve s}=\delta_{\breve{s'}}{}^{\breve s}, \cr 
\widehat Q^{(p) \breve k}=0, \qquad p\ne 2, \qquad k\ne s,
\end{eqnarray}
while others coefficients can be find from Eqs.(\ref{806}), (\ref{807}), the latters acquire the form
$(k=3, 4, \ldots ,s, ~ n=2, 3, \ldots ,s+2-k)$
\begin{eqnarray}\label{902}
\begin{array}{l}
\widehat Q^{(2)}{}_{\breve{s'}}{}^n=-Q^{(2)}{}_{\breve{s'}}{}^n, \qquad
\widehat Q^{(k)}{}_{\breve{s'}}{}^n=-Q^{(k)}{}_{\breve{s'}}{}^n-\delta_{s}{}^{k+n-2}
C_{(s) \breve{s'}}{}^n, \cr
\widehat Q^{(k)}{}_{\breve{s'}}{}^{\overline n}=\sum_{m=2}^{k-1} 
Q^{(m) k+n-m}B_{(k+n-m) k+n-m}{}^{\overline n}+ \cr 
\delta_{s}{}^{k+n-2}
\left(-D_{(s) \breve{s'}}{}^{\overline n}+\sum_{m=3}^{k-1} C_{(s) \breve{s'}}{}^{k+n-m}
B_{(k+n-m) k+n-m}{}^{\overline n}\right).
\end{array}
\end{eqnarray}
It gives a set of $s$-stage local symmetries (\ref{308}), 
number of them coincides with defect $\left[\breve{s'}\right]$ of $s$-stage Dirac system (\ref{601})  
\begin{eqnarray}\label{903}
\delta_{\breve s}q^A=
\sum^{s-2}_{p=0}\stackrel{(p)}{\epsilon}{}^{\breve s}
R^{(p)}_{\breve s}{}^A, \qquad 
R^{(p)}_{\breve s}{}^1=Q^{(s-p)}{}_{\breve s}{}^1.
\end{eqnarray}
One notes that 
\begin{eqnarray}\label{904}
\begin{array}{l}
\delta_{\breve s}q^1=
\stackrel{(s-2)}{\epsilon}{}^{\breve s}Q^{(2)}{}_{\breve s}{}^1+ \ldots=
\stackrel{(s-2)}{\epsilon}{}^{\breve s}\widetilde E_{(s) \breve s}{}^1+ \ldots , \cr 
\widetilde E_{(s) \breve s}{}^1\equiv\widetilde A_{(s) \breve s}{}^{s-1}
\widetilde A_{(s-1) s-1}{}^{s-2}\ldots\widetilde A_{(2) 2}{}^1,  
\end{array}
\end{eqnarray}
where by construction
\begin{eqnarray}\label{905}
rank\widetilde E_{(s) \breve s}{}^1=\left[\breve s\right]=\max.  
\end{eqnarray}
It implies that the symmetries obtained are independent.

Let us construct these symmetries for $s=2, 3, \ldots , N+1$. The procedure stops on the stage $s=N+1$, 
since the structure matrices $A, B, C, D$ are not defined for $N+2$. Then total number of 
the symmetries which can be constructed by using of our procedure is
\begin{eqnarray}\label{905}
\sum_{s=2}^{N+1} \left[\breve s\right]=\sum_{s=2}^{N+1} \left[s-1\right]-
\sum_{s=2}^{N+1} \left[\overline s\right]-\sum_{s=2}^{N} \left[s\right]= \cr 
\sum_{s=1}^{N} \left[s\right]-
\sum_{s=2}^{N+1} \left[\overline s\right]-\sum_{s=2}^{N} \left[s\right]= 
\left[1\right]-\sum_{s=2}^{N+1} \left[\overline s\right]=\left[v^{\underline{N+1}}\right],
\end{eqnarray}
i.e. coincides with the number of Lagrangian multipliers remaining undetermined in the Dirac 
procedure. All the symmetries obtained are independent in the sense that matrix constructed 
from the blocks $R^{(s-2)}_{\breve s}{}^1, ~ s=2, 3, \ldots , N+1$ has maximum rank by construction. 

Using Eqs.(\ref{901}), (\ref{902}), it is not difficult to write manifest form of lower-stage 
symmetries, namely 

\noindent
{\bf{Second-stage symmetries}}
\begin{eqnarray}\label{906}
\delta_{\breve 2}q^1=\epsilon^{\breve 2}\widetilde A_{(2) \breve 2}{}^1,
\end{eqnarray}
{\bf{Third-stage symmetries}}
\begin{eqnarray}\label{907}
\delta_{\breve 3}q^1=-\epsilon^{\breve 3}\left(D_{(3) \breve 3}{}^{\overline 2}
\widetilde A_{(2) \overline 2}{}^1+
C_{(3) \breve 3}{}^2\widetilde A_{(2) 2}{}^1\right)+
\dot{\epsilon}{}^{\breve 3}\widetilde A_{(3) \breve 3}{}^2\widetilde A_{(2) 2}{}^1
\end{eqnarray}
{\bf{4-stage symmetries}}
\begin{eqnarray}\label{908}
\delta_{\breve 4}q^1=-\epsilon^{\breve 4}\left(\left(D_{(4) \breve 4}{}^{\overline 2}-
C_{(4) \breve 4}{}^3B_{(3) 3}{}^{\overline 2}\right)\widetilde A_{(2) \overline 2}{}^1+
C_{(4) \breve 4}{}^2\widetilde A_{(2) 2}{}^1\right)+ \cr
\dot{\epsilon}{}^{\breve 4}\left(\widetilde A_{(4) \breve 4}{}^3B_{(3) 3}{}^{\overline 2}
\widetilde A_{(2) \overline 2}{}^1-
\left(D_{(4) \breve 4}{}^{\overline 3}\widetilde A_{(3) \overline 3}{}^2+
C_{(4) \breve 4}{}^3\widetilde A_{(3) 3}{}^2\right)\widetilde A_{(2) 2}{}^1\right)+ \cr
\ddot{\epsilon}{}^{\breve 4}\widetilde A_{(4) \breve 4}{}^3\widetilde A_{(3) \breve 3}{}^2
\widetilde A_{(2) 2}{}^1.
\end{eqnarray}
Thus knowledge of structure matrices $A, B, C, D$ of the Dirac procedure determines independent 
local symmetries of the Lagrangian action.  Number of the symmetries coincides with number of 
Lagrangian multipliers remaining arbitrary in the end of Dirac procedure. 
Surprising conclusion following from the present analysis is that search for gauge 
symmetries in quadratic theory do not requires separation of the Dirac constraints 
on first and second class subsets. 

\section{Acknowledgments}
Author would like to thank the Brazilian foundations CNPq and FAPEMIG 
for financial support.

\section{Appendix A}

We present here three Lemmas which are used in section 5 to rewrite $p$-stage Dirac functions in 
the normal form (\ref{601}).

Suppose $\triangle_{p-1, \underline{p-1}}(z_1)$ be a matrix with $rank\triangle=\left[\overline p\right], ~ 
\left[p-1\right]-\left[\overline p\right]=\left[\widetilde p\right]$. Let
$K_{\widetilde{p}}{}^{p-1}$ represents complete set of independent null-vectors of $\triangle$ and 
$K_{\overline{p}}{}^{p-1}$ be any completion of the set 
$\vec K_{\widetilde{p}}$ up to base of $[p-1]$-dimensional space. By construction, the matrix    
\begin{eqnarray}\label{2.1}
K_{\widehat{p-1}}{}^{p-1}=
\left(K_{\overline{p}}{}^{p-1}\atop  K_{\widetilde{p}}{}^{p-1}\right),
\end{eqnarray}
is invertible, with the inverse matrix being 
$\widetilde K_{p-1}{}^{\widehat{p-1}} K_{\widehat{p-1}}{}^{p-1'}=\delta_{p-1}{}^{p-1'}$. 

\noindent
{\bf{Lemma 1}} For some set of variables $v^{\underline{p-1}}$, let us consider linear functions 
$\triangle_{p-1 \underline{p-1}}v^{\underline{p-1}}+L_{p-1}(z_1)+M_{p-1}(z_1, \Pi)$, where $\Pi$ is 
group of some variables. Then\par
\noindent
a) The functions  can be identically rewritten in the form
\begin{eqnarray}\label{2.2}
\triangle_{p-1, \underline{p-1}}v^{\underline{p-1}}+L_{p-1}+M_{p-1}=
\widetilde K_{p-1}{}^{\widehat{p-1}}\left(\pi_{\overline{p}}(v^{\overline{p-1}})\atop 
\Phi_{\widetilde{p}}(z_1)+N_{\widetilde p}(z_1, \Pi)\right),
\end{eqnarray}
\begin{eqnarray}\label{2.3}
\pi_{\overline{p}}\equiv X_{\overline p \underline{p-1}}v^{\underline{p-1}}+Y_{\overline p}, \cr 
\Phi_{\widetilde{p}}\equiv K_{\widetilde p}{}^{p-1} L_{p-1}, \cr 
N_{\widetilde p}\equiv K_{\widetilde p}{}^{p-1}M_{p-1}.
\end{eqnarray}
Here it was denoted 
\begin{eqnarray}\label{2.4}
X_{\overline p \underline{p-1}}=K_{\overline p}{}^{p-1}
\triangle_{p-1, \underline{p-1}},\qquad
rank X=\left[\pi_{\overline p}\right], \cr 
Y_{\overline p}=K_{\overline p}{}^{p-1}L_{p-1}. \qquad \qquad \qquad \qquad \qquad \qquad  
\end{eqnarray}
b) $v^{\underline{p-1}}$ can be divided on groups $v^{\overline{p}}, v^{\underline{p}}$ 
in such a way that there is identity 
\begin{eqnarray}\label{2.5}
v^{\overline{p}}\equiv\widetilde X^{\overline{p} \overline{p'}}
\pi_{\overline{p'}}(v^{\underline{p-1}})+
\Lambda^{\overline{p}}{}_{\underline{p}}v^{\underline{p}}+W^{\overline{p}}.
\end{eqnarray}

{\bf{Proof.}} Computing product of $1=\widetilde K K$ with $\triangle v+L+M$ one obtains 
Eqs.(\ref{2.2})-(\ref{2.4}). Let us show that $rank X=\left[\pi_{\overline p}\right]$. 
Suppose that $rank X<\left[\pi_{\overline p}\right]$, then there is null-vector 
$\xi^{\overline p}
K_{\overline p}{}^{p-1}\triangle_{p-1 \underline{p-1}}=0$. If 
$\xi^{\overline p}K_{\overline p}{}^{p-1}=0$, the vectors $\vec K_{\overline p}$ are 
linearly dependent, which is contradiction. If 
$\left[\xi^{\overline p}
K_{\overline p}{}^{p-1}\right]\triangle_{p-1 \underline{p-1}}=0$, 
the linear combination $\xi^{\overline p}\vec K_{\overline p}$ of the vectors $\vec K_{\overline p}$ 
is null vector of $\triangle$. It is impossible, since $\vec K_{\widetilde p}$ form basis of 
null-vectors and $\vec K_{\overline p}$ are linearly independent of them by construction.

Since $rank X=\left[\pi_{\overline p}\right]=\max$, the first equation in (\ref{2.3}) can be resolved in 
relation of some subgroup $v^{\overline p}$ of the group 
$v^{\underline{p-1}}=(v^{\overline p}, v^{\underline p})
$, which implies the identity Eq.(\ref{2.5}). $\spadesuit$

Suppose a group $v^1$ is divided on subgroups as follows: 
$v^1=(v^{\overline{2}}, v^{\underline{2}})$, 
$v^{\underline{2}}=(v^{\overline{3}}, v^{\underline{3}}), \ldots , 
v^{\underline{p-2}}=(v^{\overline{p-1}}, v^{\underline{p-1}})$, with the resulting division being
$v^1=(v^{\overline{2}}, v^{\overline{3}}, \ldots ,v^{\overline{p-1}}, 
v^{\underline{p-1}})$. Consider the functions 
\begin{eqnarray}\label{2.6}
\pi_{\overline{n}}(v^{\underline{n-1}})=
X_{(n) \overline n \underline{n-1}}v^{\underline{n-1}}+Y_{(n) \overline n}, \cr
rank X=\left[\pi_{\overline n}\right], ~ n=2, 3, \ldots ,p-1.
\end{eqnarray}
As it was discussed in the proof of the Lemma 1, one writes the identities
\begin{eqnarray}\label{2.7}
v^{\overline{n}}\equiv\widetilde X_{(n)}{}^{\overline{n} \overline{n'}}
\pi_{\overline{n'}}(v^{\underline{n-1}})+
\Lambda_{(n)}{}^{\overline{n}}{}_{\underline{n}}v^{\underline{n}}+W_{(n)}{}^{\overline{n}}.
\end{eqnarray}

\noindent
{\bf{Lemma 2}} Let $3\le k\le p$ be some fixed number and $\nabla_{p-1, 1}(z_1)$, $H_{p-1}(z_1)$ 
are some quantities. Then there is identity 
\begin{eqnarray}\label{2.8}
\nabla_{p-1, 1}v^1+H_{p-1}= \qquad \qquad \qquad \cr  
\triangle_{(k)p-1, \underline{k-1}}v^{\underline{k-1}}+L_{(k) p-1}+
M_{(k) p-1}(\pi_{\overline{k-1}}, \ldots ,\pi_{\overline 2}), 
\end{eqnarray}
where the quantities $\triangle, ~ L, ~ M$ can be find from the following recurrence relations  
(for $k=2$ one takes $\triangle_{(2)p-1,1}=\nabla_{p-1,1}$, $L_{(2)p-1}=H_{p-1}$, 
$M_{(2)p-1}=0$): 
\begin{eqnarray}\label{2.9}
\triangle_{(k)p-1, \underline{k-1}}=\triangle_{(k-1)p-1, \underline{k-1}}+
\triangle_{(k-1)p-1 \overline{k-1}}\Lambda_{(k-1)}{}^{\overline{k-1}}{}_{\underline{k-1}}, \cr
L_{(k)p-1}=L_{(k-1)p-1}+\triangle_{(k-1)p-1, \overline{k-1}}+W_{(k-1)}{}^{\overline{k-1}}, \qquad \quad \cr 
M_{(k)p-1}=M_{(k-1)p-1}+\triangle_{(k-1)p-1, \underline{k-1}}
\widetilde X_{(k-1)}{}^{\overline{k-1}, \overline{k-1'}}\pi_{\overline{k-1'}}.
\end{eqnarray}

\noindent
{\bf{Proof.}} (Induction over $k$). For $k=3$ one has 
\begin{eqnarray}\label{2.10}
\begin{array}{l}
\nabla_{p-1, 1}v^1+H_{p-1}=\nabla_{p-1, \underline 2}v^{\underline 2}+ 
\nabla_{p-1, \overline 2}
\left(\widetilde X_{(2)}{}^{\overline{2} \overline{2'}}\pi_{\overline{2'}}+ 
\Lambda_{(2)}{}^{\overline{2}}{}_{\underline{2}}v^{\underline{2}}+ \right. \cr
\left. W_{(2)}{}^{\overline{2}}\right)H_{p-1}=
\left(\nabla_{p-1, \underline 2}+\nabla_{p-1, \overline 2}
\Lambda_{(2)}{}^{\overline{2}}{}_{\underline{2}}\right)
v^{\underline 2}+\left(H_{p-1}+ 
\nabla_{p-1, \overline 2}W_{(2)}{}^{\overline{2}}\right)+ \cr
\nabla_{p-1, \overline 2}\widetilde X_{(2)}{}^{\overline{2} \overline{2'}}\pi_{\overline{2'}}\equiv 
\triangle_{(3)p-1, \underline{2}}v^{\underline{2}}+L_{(3) p-1}+
M_{(3) p-1}(\pi_{\overline{2}}),
\end{array}
\end{eqnarray}
in accordance with (\ref{2.9}). Now, let us suppose that the Lemma is true for $k-1$. According to the 
induction hypothesis, one writes
\begin{eqnarray}\label{2.11}
\nabla_{p-1, 1}v^1+H_{p-1}= \qquad \qquad \qquad \cr \nonumber 
\triangle_{(k-1)p-1, \underline{k-2}}v^{\underline{k-2}}+L_{(k-1) p-1}+
M_{(k-1) p-1}. 
\end{eqnarray}
Substitution of $v^{\underline{k-2}}$ in the form $v^{\overline{k-1}}, v^{\underline{k-1}}$, where 
$v^{\overline{k-1}}$ is given by Eq.{\ref{2.7}) leads, similarly to the previous computation, to
Eqs.(\ref{2.8}), (\ref{2.9}). $\spadesuit$

Let $\Phi(z_1)$ be system of functionally independent functions,  and $z_1$ is divided on 
$(\bar z_2, z_2)$ in such a way that 
\begin{eqnarray}\label{2.12}
rank\frac{\partial\Phi}{\partial z_1}\Biggr|_{\Phi}=rank\frac{\partial\Phi}{\partial\bar z_2}\Biggr|_{\Phi}=
\left[\Phi\right]=\left[\bar z_2\right].
\end{eqnarray}
Let $\Phi_{\widetilde p}(z_1)$ are some functions and 
\begin{eqnarray}\label{2.13}
rank\frac{\partial(\Phi, \Phi_{\widetilde p})}{\partial z_1}\Biggr|_{\Phi, \Phi_{\widetilde p}}=a.
\end{eqnarray}

\noindent
{\bf{Lemma 3.}} There is the representation
\begin{eqnarray}\label{2.14}
\Phi_{\widetilde{p}}(z_1)=U_{\widetilde{p}}{}^{\widetilde{p}}
\left(\Phi_p(z_1)\atop C_{\breve p}(\Phi)\right), 
\end{eqnarray}
where\par
\noindent
a)$U(z_1)$ is invertible matrix. \par
\noindent
b) $C_{\breve p}(\Phi)$ are linear homogeneous functions with the coefficients dependent on $z_1$ only.\par
\noindent
c) $z_2=(\bar z_3, z_3)$ such that 
\begin{eqnarray}\label{2.155}
rank\frac{\partial\Phi_p}{\partial z_1}\Biggr|_{\Phi_p}= 
rank\frac{\partial\Phi_p}{\partial\bar z_3}\Biggr|_{\Phi_p}
=\left[\Phi_p\right]=\left[\bar z_3\right]=a-\left[\Phi\right].
\end{eqnarray}
d) Functions $\Phi, ~ \Phi_p$ are functionally independent.

This result means that the system of independent functions $\Phi, ~ \Phi_p$ is equivalent to the 
initial system $\Phi, ~ \Phi_{\widetilde p}$.

\noindent
{\bf{Proof.}} Eqs.(\ref{2.12}), (\ref{2.13}) imply that the matrix  
$\frac{\partial(\Phi, \Phi_{\widetilde p})}{\partial z_1}$ has 
rank minor composed of lines $\Phi$ and some of lines of $\Phi_{\widetilde p}$. One uses numeric invertible 
matrix $Q$ to rearrange the lines 
\begin{eqnarray}\label{2.15}
\left(\Phi\atop \Phi_{\widetilde p}\right)=
\left(
\begin{array}{cc}
1_{[\Phi]\times[\Phi]}&0 \\
0&Q_{[\widetilde p]\times [\widetilde p]}
\end{array}
\right)
\left(
\begin{array}{c}
\Phi\\
\Phi '_{\widetilde p}\\
\Phi ''_{\widetilde p}\\
\end{array}
\right)
\end{eqnarray}
in such a way that $(\Phi, \Phi '_{\widetilde p})$ are functionally independent  
\begin{eqnarray}\label{2.16}
rank\frac{\partial(\Phi , \Phi '_{\widetilde p})}{\partial z_1}
\Biggr|_{\Phi, \Phi_{\widetilde p}}=a, \quad
rank\frac{\partial\Phi '_{\widetilde p}}{\partial z_1}
\Biggr|_{\Phi, \Phi_{\widetilde p}}= 
\left[\Phi '_{\widetilde p}\right]=a-\left[\Phi\right].
\end{eqnarray}
Under these conditions the right column in Eq.(\ref{2.15}) has the representation [11]
\begin{eqnarray}\label{2.17}
\left(
\begin{array}{c}
\Phi\\
\Phi '_{\widetilde p}\\
\Phi ''_{\widetilde p}\\
\end{array}
\right)=
\left(
\begin{array}{cc}
E&F\\
G&H\\
M&N
\end{array}
\right)
\left(
\begin{array}{c}
\Phi '(\bar z_2, z_2)\\
\Phi '_p(z_2)\\
\end{array}
\right)
\end{eqnarray}
where ($S$ is the block matrix)  
\begin{eqnarray}\label{2.18}
rank S|_{\Phi , \Phi_{\widetilde p}}=a, \qquad 
rank\frac{\partial\Phi '}{\partial z_1}\Biggr|_{\Phi '}=
rank\frac{\partial\Phi}{\partial \bar z_2}\Biggr|_{\Phi}=
\left[\Phi '\right]= \cr
\left[\Phi\right]=\left[\bar z_2\right] \qquad 
rank\frac{\partial\Phi '_p}{\partial z_2}\Biggr|_{\Phi '_p}
=\left[\Phi '_p\right]=a-\left[\Phi\right].
\end{eqnarray}
According to these conditions, the functions $\Phi '(\bar z_2, z_2), \Phi '_p(z_2)$ are functionally independent, 
and $z_2$ can be divided on $(\bar z_3, z_3)$, $\left[\bar z_3\right]=\left[\Phi '_p\right]$, such that  
$\bar z_2, \bar z_3$ can be find from equations $\Phi '(\bar z_2, z_2)=0, ~ \Phi '_p(z_2)=0$ in terms of 
$z_3$. 

Since $(\Phi, \Phi '_{\widetilde p})$, $(\Phi ', \Phi '_p)$ and $\Phi$
are systems of functionally independent functions, 
one has
{\footnote{a) Let $\phi=K\psi$ with functionally independent sets $\phi$ and $\psi$, and 
$\left[\phi\right]=\left[\psi\right]$. Suppose $\det K=0$, then one uses null-vectors of $K$ 
(similarly to section 4) to write
$\phi=Q\left(\psi '\atop 0\right), ~  \det Q\ne 0$, which is in contradiction with 
functionally independence of $\phi$.\par
\noindent
b) From the first line of Eq.(\ref{2.17}) one has $\Phi=E\Phi '+F\Phi '_p$, then 
$0\ne\det\frac{\partial\Phi}{\partial\bar z_2}|_{\Phi}=
\det E\det\frac{\partial\Phi '}{\partial\bar z_2}|_{\Phi ', \Phi '_p}$, then $\det E\ne 0$.}}
\begin{eqnarray}\label{2.19}
\det
\left(
\begin{array}{cc}
E&F\\
G&H
\end{array}
\right)
\ne 0, \qquad \qquad
\det E\ne 0.
\end{eqnarray}
It allows one to write Eq.(\ref{2.17}) in the form
\begin{eqnarray}\label{2.20}
\left(
\begin{array}{c}
\Phi\\
\Phi '_{\widetilde p}\\
\Phi ''_{\widetilde p}\\
\end{array}
\right)=
\left(
\begin{array}{cc}
1&0\\
\Lambda_1&\Lambda_2\\
\Lambda_3&\Lambda_4
\end{array}
\right)
\left(
\begin{array}{c}
E\Phi '+F\Phi '_p\\
\Phi '_p\\
\end{array}
\right).
\end{eqnarray}
where $\Lambda_1=G \widetilde E, ~ \Lambda_2=H-G \widetilde E F, ~ \Lambda_3=M \widetilde E, ~ 
\Lambda_4=N-M \widetilde E F$, 
and $\Lambda_2$ is invertible. One concludes $E\Phi '+F\Phi '_p=\Phi$, then Eq.(\ref{2.20}) implies 
\begin{eqnarray}\label{2.21}
\left(
\begin{array}{c}
\Phi '_{\widetilde p}\\
\Phi ''_{\widetilde p}\\
\end{array}
\right)=
\left(
\begin{array}{cc}
\Lambda_2&0\\
\Lambda_4&1
\end{array}
\right)
\left(
\begin{array}{c}
\Phi '_p+\widetilde\Lambda_2\Lambda_1\Phi\\
\left(\Lambda_3-\Lambda_4\widetilde\Lambda_2\Lambda_1\right)\Phi\\
\end{array}
\right).
\end{eqnarray}
Let us denote 
\begin{eqnarray}\label{2.22}
\Phi_p\equiv\Phi '_p+\widetilde\Lambda_2\Lambda_1\Phi, \qquad
C\equiv\Lambda_3-\Lambda_4\widetilde\Lambda_2\Lambda_1.
\end{eqnarray}
Since $\Phi '_{\widetilde p}=\Lambda_2\Phi_p$ with $\Lambda_2$ invertible, the systems are equivalent, 
and 
$rank\frac{\partial\Phi_p}{\partial z_1}\Biggr|_{\Phi_p}=\left[\Phi_p\right]=
\left[\Phi '_{\widetilde p}\right]=
a-\left[\Phi\right]$. By construction, the $(\Phi, \Phi_p)$ are functionally independent. Combining 
Eqs.(\ref{2.15}), (\ref{2.21}) one finds the desired result (\ref{2.14}), where 
$U=Q
\left(
\begin{array}{cc}
\Lambda_2&0\\
\Lambda_4&1
\end{array}
\right)$ is invertible.

\section{Appendix B}

We present here basic Lemma which was used in section 8 to rewrite $s$-stage generating functions in 
the normal form (\ref{802}).

{\bf{Lemma 4}}. Consider an expression
\begin{eqnarray}\label{3.1}
T^{(p)}=Q^1\{\Phi_1, H\}-\{H, \sum_{n=2}^{p-1} Q^n\Phi_n\}, 
\end{eqnarray}
with arbitrary functions $Q^n(q^A, p_j), ~ n=1, 2, \ldots ,p-1$.
Then $Q^n$ can be chosen in such a way, that $T^{(p)}$ will be linear combinations of the constraints  
\begin{eqnarray}\label{3.2}
T^{(p)}=-\sum_{n=2}^p Q'^n\Phi_n, 
\end{eqnarray}
where $Q'^n$ are arbitrary functions. Choice of $Q^n$, which supplies the normal form can be described 
as follow: \par
\noindent
a) For any $n$, ~  $Q^n$ is divided on three subgroups with help of the structure matrix $A_{(n+1)}$ 
(see Eq.(\ref{601}))  
\begin{eqnarray}\label{3.3}
Q^nA_{(n+1) n}{}^{\widehat n}\equiv\widehat Q^{\widehat n}=
(\widehat Q^{\overline{n+1}}, \widehat Q^{n+1}, \widehat Q^{\breve{n+1}}), 
\end{eqnarray}
where for any $n=2,3, \ldots ,p$ one has 
\begin{eqnarray}\label{3.4}
\widehat Q^n=-Q'^n+\{H, \widehat Q^{\widehat n}
\widetilde A_{(n+1) \widehat n}{}^n\}- \qquad \cr 
\sum_{m=n}^{p-1} \widehat Q^{\breve{m+1}}C_{(m+1) \breve{m+1}}{}^n, \qquad 
\Longrightarrow \widehat Q^{p}=-Q'^{p},
\end{eqnarray}
\begin{eqnarray}\label{3.5}
\widehat Q^{\overline n}=-
\sum_{m=n}^{p-1}\left(\widehat Q^{m+1}B_{(m+1) m+1}{}^{\overline n}+\right. \cr
\left. \widehat Q^{\breve{m+1}}D_{(m+1) \breve{m+1}}{}^{\overline n}\right) \qquad \quad 
\Longrightarrow \widehat Q^{\overline p}=0,
\end{eqnarray}
and $B, ~ C, ~ D$ are structure matrix of Eq.(\ref{601}). \par
\noindent
b) The coefficients $\widehat Q^{\breve n}, ~ n=2,3, \ldots ,p$ remains 
arbitrary. 

\noindent
{\bf Proof.}} Using normal form of the Dirac functions given by Eq.(\ref{601}), one obtains the following 
expression for $T^{(p)}$ 
\begin{eqnarray}\label{3.6}
T^{(p)}=\widehat Q^{\widehat1}
\left(
\begin{array}{c}
\pi_{\overline{2}}\\
\Phi_2\\
0_{\breve 2}
\end{array}
\right)-\sum_{n=2}^{p-1}\{H, Q^n\}\Phi_n+ \qquad \qquad \\[3pt] \nonumber
\sum_{n=2}^{p-1}\widehat Q^{\widehat n}
\left(
\begin{array}{c}
\pi_{\overline{n+1}}\\
\Phi_{n+1}+B_{(n+1) n+1}(\pi_{\overline{n}}, \ldots ,\pi_{\overline{2}})\\
C_{(n+1)\breve{n+1}}(\Phi_n, \ldots ,\Phi_2)+
D_{(n+1)\breve{n+1}}(\pi_{\overline n}, \ldots ,\pi_{\overline 2})
\end{array}
\right)
\end{eqnarray}
where $\widehat Q$ are given by Eq.(\ref{3.3}). Collecting similar terms in Eq.(\ref{3.6}) and using 
the identity 
\begin{eqnarray}\label{3.7}
\sum_{n=2}^{p-1} \sum_{m=2}^n K(n)^{\overline m}\pi_{\overline m}=
\sum_{n=2}^{p-1} \left(\sum_{m=n}^{p-1} K(m)^{\overline n}\right)\pi_{\overline n},
\end{eqnarray}
one obtains
\begin{eqnarray}\label{3.8}
T^{(p)}= \qquad \qquad \qquad \qquad \\[1pt] \nonumber  
\sum_{n=2}^{p-1}
\left(\widehat Q^{\overline n}+
\sum_{m=n}^{p-1}
\left(\widehat Q^{m+1}B_{(m+1) m+1}{}^{\overline n}+
\widehat Q^{\breve{m+1}}D_{(m+1)\breve{m+1}}{}^{\overline n}\right)\right)\pi_{\overline{n}}+ \cr 
\widehat Q^{\overline p}\pi_{\overline{p}}+ \qquad \qquad \qquad \qquad \\[1pt] \nonumber
\sum_{n=2}^{p-1}
\left(\widehat Q^{n}-\{H, Q^{n}\}+\sum_{m=n}^{p-1}\widehat Q^{\breve{m+1}}
C_{(m+1)\breve{m+1}}{}^n\right)\Phi_n+ \cr 
\widehat Q^{p}\Phi_p+
\widehat Q^{\breve 2}0_{\breve 2}.\qquad \qquad \qquad 
\end{eqnarray}
Then the choice of $Q$ written in Eqs.(\ref{3.4}), (\ref{3.5}) leads to the normal form  
(\ref{3.2}) for $T^{(p)}$. $\spadesuit$

Evolution of the functions $Q$ of Eq.(\ref{3.1}) can be described schematically as follows:
\begin{eqnarray}\label{3.9}
\left(
\begin{array}{c}
Q^{p-1}\\
\ldots\\
Q^2\\
Q^1\\
\end{array}
\right)\longrightarrow
\left(
\begin{array}{ccc}
\widehat Q^{\overline p}&\widehat Q^p&\widehat Q^{\breve p}\\
\ldots&\ldots&\ldots\\
\widehat Q^{\overline 3}&\widehat Q^3&\widehat Q^{\breve 3}\\
\widehat Q^{\overline 2}&\widehat Q^2&\widehat Q^{\breve 2}\\
\end{array}
\right)\longrightarrow
\left(
\begin{array}{ccc}
\widehat Q^{\overline p}&Q'^p&\widehat Q^{\breve p}\\
\ldots&\ldots&\ldots\\
\widehat Q^{\overline 3}&Q'^3&\widehat Q^{\breve 3}\\
0&Q'^2&\widehat Q^{\breve 2}\\
\end{array}
\right)
\end{eqnarray}
Being arranged in this order, the lines correspond to the ones in Eq.(\ref{805}), see Eq.(\ref{811}). 
Functions placed in second and third columns of the last matrix remain arbitrary. 
According to Eq.(\ref{3.5}), any group $\widehat Q^{\overline n}$ of the line $n$ is presented through 
functions of previous lines placed in second and third columns. According to Eq.(\ref{3.4}), any 
$\widehat Q^n$ is presented in terms of arbitrary function $Q'^{n}$ as well as through functions of 
previous lines. 
  
It is convenient to write manifest form for division of the coefficients $Q^{(p) 1}$ on $s$-stage of 
the Dirac procedure, namely
\begin{eqnarray}\label{3.10}
\begin{array}{l}
Q^{(2) 1}=\left(0^{\overline 2}, \left(0^{\overline 3}, \left(\ldots , \left(0^{\overline{s-1}},
\left(0^{\overline{s}}, \widehat Q^{(2) s} 
\widehat Q^{(2) \breve s}\right)^{\widehat{s-1}}\right.\right.\right.\right. \cr
\left.\left.\left.\left.
\widetilde A_{(s) \widehat{s-1}}{}^{s-1}, 
\widehat Q^{(2) \breve{s-1}}\right)^{\widehat{s-2}}\widetilde A_{(s-1) \widehat{s-2}}{}^{s-2},
\widehat Q^{(2) \breve{s-2}}\right)^{\widehat{s-3}}\ldots , 
\widehat Q^{(2) \breve 3}\right)^{\widehat 2}\right. \cr 
\left.\widetilde A_{(3) \widehat 2}{}^2, 
\widehat Q^{(2) \breve 2}\right)^{\widehat 1}\widetilde A_{(2) \widehat 1}{}^1, \cr 
Q^{(3) 1}=\left(\widehat Q^{(3) \overline 2}, \left(\ldots\left(\widehat Q^{(3) \overline{s-2}}, 
\left(\widehat Q^{(3) \overline{s-1}},
\widehat Q^{(3) s-1} 
\widehat Q^{(3) \breve{s-1}}\right)^{\widehat{s-2}}\right.\right.\right. \cr
\left.\left.\left.\widetilde A_{(s-1) \widehat{s-2}}{}^{s-2}, 
\widehat Q^{(3) \breve{s-2}}\right)^{\widehat{s-3}}\widetilde A_{(s-2) \widehat{s-3}}{}^{s-3},
\widehat Q^{(3) \breve{s-3}}\right)^{\widehat{s-4}}\ldots , 
\widehat Q^{(3) \breve 2}\right)^{\widehat 1}\widetilde A_{(2) \widehat 1}{}^1, \cr
\ldots \qquad \qquad \qquad \ldots \qquad \qquad \qquad \ldots \qquad \qquad \qquad \ldots \cr
Q^{(p) 1}=\left(\widehat Q^{(p) \overline 2}, \left(\left.\ldots\left(\widehat Q^{(p) \overline{s+2-p}}, 
\widehat Q^{(p) s+2-p} 
\widehat Q^{(p) \breve{s+2-p}}\right)^{\widehat{s+1-p}}\right.\right.\right. \cr
\left.\left.\left.\widetilde A_{(s+2-p) \widehat{s+1-p}}{}^{s+1-p}, 
\widehat Q^{(p) \breve{s+1-p}}\right)^{\widehat{s-p}}\widetilde A_{(s+1-p) \widehat{s-p}}{}^{s-p},
\widehat Q^{(p) \breve{s-p}}\right)^{\widehat{s-p-1}}\ldots , \right. \cr  
\left.\widehat Q^{(p) \breve 2}\right)^{\widehat 1}\widetilde A_{(2) \widehat 1}{}^1, \cr
\ldots \qquad \qquad \qquad \ldots \qquad \qquad \qquad \ldots \qquad \qquad \qquad \ldots \cr
\end{array}
\cr
\begin{array}{l}
Q^{(s-1) 1}=\left(\widehat Q^{(s-1) \overline 2}, \left(\widehat Q^{(s-1) \overline 3}, 
\widehat Q^{(s-1) 3} 
\widehat Q^{(s-1) \breve 3}\right)^{\widehat 2}\widetilde A_{(3) \widehat 2}{}^2,
\widehat Q^{(s-1) \breve 2}\right)^{\widehat 1}\widetilde A_{(2) \widehat 1}{}^1, \cr
Q^{(s) 1}=\left(\widehat Q^{(s) \overline 2}, 
\widehat Q^{(s) 2},  
\widehat Q^{(s) \breve 2}\right)^{\widehat 1}\widetilde A_{(2) \widehat 1}{}^1,
\end{array}
\end{eqnarray}

\end{document}